\documentclass[aps,prb,twocolumn,preprintnumbers,superscriptaddress,amsmath]{revtex4}

\newcommand{\suppinfo}{Supplemental Materials~\cite{supp-info}}
\usepackage{amsmath}
\usepackage{amssymb}
\usepackage{enumitem}
\usepackage{physics}
\usepackage[dvipsnames]{xcolor}
\usepackage[T1]{fontenc}
\usepackage{graphicx}
\usepackage{dcolumn}
\usepackage{bm}
\usepackage[retainorgcmds]{IEEEtrantools}
\usepackage{xcolor}
\usepackage[colorlinks=true,linkcolor=blue,citecolor=blue,urlcolor=blue]{hyperref}
\usepackage{ulem}
\usepackage{bbold}
\usepackage{mathtools}
\usepackage{accents}
\usepackage{verbatim} % comments

\newcommand{\yambo}{\textsc{yambo}}

\newcommand{\hide}[1]{}

\newcommand{\SE}{\Sigma}
\newcommand{\SEx}{\Sigma^{\mathrm{x}}}
\newcommand{\SEc}{\Sigma^{\mathrm{c}}}
\newcommand{\scr}{W}
\newcommand{\scrc}{W^{c}}

\newcommand{\G}{\mathbf{G}}
\newcommand{\q}{\mathbf{q}}

\renewcommand{\o}{\omega}

\newcommand{\eig}{\varepsilon}

\renewcommand{\k}{\mathbf{k}}
\newcommand{\dm}{\rho}
\newcommand{\inveps}{\epsilon^{-1}}

\newcommand{\vcoul}{v}
\newcommand{\sgn}{\mathrm{sgn}}
\newcommand{\Rp}{R}
\newcommand{\Op}{\Omega}

\renewcommand{\emph}{\textit}

\begin{document}

\title
{Efficient GW calculations via the interpolation of the screened interaction in momentum and frequency space: The case of graphene}

\author{Alberto Guandalini}
\email{alberto.guandalini@uniroma1.it}
\altaffiliation[Current Address: ]{Dipartimento di Fisica, Universit\`a di Roma La Sapienza, Piazzale Aldo Moro 5, I-00185 Roma, Italy}
\affiliation{S3 Centre, Istituto Nanoscienze, CNR, Via Campi 213/a, Modena (Italy)}
\thanks{A. Guandalini and D. A. Leon contributed equally to this work.}

\author{Dario A. Leon}
\email{dario.alejandro.leon.valido@nmbu.no}
\affiliation{S3 Centre, Istituto Nanoscienze, CNR, Via Campi 213/a, Modena (Italy)}
\affiliation{Department of Mechanical Engineering and Technology Management, Norwegian University of Life Sciences, 1430, Ås (Norway)}
\thanks{A. Guandalini and D. A. Leon contributed equally to this work.}

\author{Pino D'Amico}
\affiliation{S3 Centre, Istituto Nanoscienze, CNR, Via Campi 213/a, Modena (Italy)}
\author{Claudia Cardoso}
\affiliation{S3 Centre, Istituto Nanoscienze, CNR, Via Campi 213/a, Modena (Italy)}
\author{Andrea Ferretti}
\affiliation{S3 Centre, Istituto Nanoscienze, CNR, Via Campi 213/a, Modena (Italy)}
\author{Massimo Rontani}
\affiliation{S3 Centre, Istituto Nanoscienze, CNR, Via Campi 213/a, Modena (Italy)}
\author{Daniele Varsano}
\affiliation{S3 Centre, Istituto Nanoscienze, CNR, Via Campi 213/a, Modena (Italy)}

\date{ \today}

\begin{abstract}
The $GW$ self-energy may become computationally challenging to evaluate because of frequency and momentum convolutions.
These difficulties were recently addressed by the development of the 
multipole approximation (MPA) and the W-av methods:
MPA accurately approximates full-frequency response functions using a small number of poles, while  W-av improves the convergence with respect to the $\mathbf{k}$-point sampling in 2D materials.
In this work we ($i$) present a theoretical scheme to combine them, and ($ii$) apply the newly developed approach to the paradigmatic case of graphene. Our findings show an excellent agreement of the calculated QP band structure with angle resolved photoemission spectroscopy (ARPES) data.
Furthermore, the computational efficiency of MPA and W-av allows us to explore the logarithmic renormalization of the Dirac cone. To this aim, we develop an analytical model, derived from a Dirac Hamiltonian, that we parameterize using {\it ab-initio} data.  
The comparison of the models obtained with PPA and  MPA results highlights an important role of the dynamical screening in the cone renormalization. 
\end{abstract}

\maketitle

\section{Introduction}
State of the art many-body perturbation theory from first principles provides accurate prediction of electronic excitations in materials.~\cite{reining2018gw,Golze_2019,Marzari2021NatMaterials}
In particular, the $GW$ approximation~\cite{Hedin_1965,Strinati_1982,Hybertsen_1986,Godby_1988} gives access to properties associated with charged excitations, such as quasi-particle band structures,~\cite{Damascelli2003RMP} satellites,~\cite{Guzzo2011PRL,Caruso2018PRB} lifetimes,~\cite{Marini2002PRB} and spectral functions.~\cite{Bechstedt1994PRB,Gatti2020PNAS}
The $GW$ approximation has been applied to a variety of materials, ranging from solids, to molecules and nanostructures.~\cite{reining2018gw,Golze_2019}
Among these, several studies~\cite{Trevisanutto_2008,haastrup2018computational,rasmussen2021towards,qian2014quantum,varsano2020monolayer,qiu2016screening,Qiu_2017,DaJornada_2020,Yabei_2018,Zhang_2019,Qiu_2020,bonacci2022excitonic} have focused on two-dimensional (2D) materials, due to their potential application in  optoelectronics, sensing, mechanical and energy storage technologies.~\cite{Ferrari_2015,Bhimanapati_2015} 

The $GW$ approximation is usually implemented in its non-self-consistent framework, namely the $G_0W_0$ approximation, where the Green's function, $G_0$, and the screened interaction, $W_0$, are calculated in one shot from DFT energies and wavefunctions.
Already without self-consistency, the computation of quasi-particle (QP) properties within $G_0W_0$ often requires a significant computational effort, in particular in the evaluation of convolution integrals between $G_0$ and $W_0$, both in frequency and transferred momentum.
Many implementations of the $GW$ method solve the frequency integral in the self-energy resorting to the plasmon-pole approximation (PPA),~\cite{Hybertsen_1986,Zhang_1989,Godby_1989,Linden_1988,Engel_1993} where the frequency dependence of the screened interaction is simplified through an analytical model with a single pole for each matrix element.
Within PPA, the screened interaction is calculated for only one or two frequency values, and the frequency integral is computed analytically, making the calculation more efficient with respect to full-frequency approaches.
However, the accuracy of PPA is strongly system dependent.~\cite{Valido_2021,leon2023efficient}

Several full-frequency (FF) methods have been proposed,~\cite{Kresse_2006,Chang_1994,Marini_2001,Liu_2015,Miyake_2000} most of them making use of numerical approaches to integrate the self-energy by means of quadrature rules, the  contour deformation (CD) technique,~\cite{Godby_1988,kotani2007quasiparticle,book_Anisimov2000,govoni2015large,farid1988gw,lebegue2003implementation} or analytic continuation methods.~\cite{rojas1995space,Riegera1999CPC,Liu2016PRB,wilhelm2018toward,duchemin2020robust}
Recently, a multipole approximation (MPA) for the frequency dependence of the screened potential has been developed.~\cite{Valido_2021,leon2023efficient}
As in PPA, the self-energy is integrated analytically, with a computational cost similar to PPA, but an accuracy closer to FF methods.
In fact, MPA achieves full-frequency quality results by evaluating the screened interaction at tens of points in the complex frequency space, instead of hundreds/thousands as for example in real-axis approaches  (FF-RA).~\cite{Valido_2021,leon2023efficient}

Low dimensional systems, like 2D systems, pose specific challenges to the calculation of the momentum transfer ($\mathbf{q}$) integrals in the self-energy. 
%\AGcancel{In the long-wavelength limit\DAL{, $\mathbf{q}\to 0$,}}
At low momentum transfers,
the dielectric function may vary very rapidly with $\mathbf{q}$, as seen e.g. for semiconductors,~\cite{qiu2016screening,Huser_2013} which makes the integration over the Brillouin zone (BZ) computationally expensive. 
In Ref.~[\onlinecite{Guandalini2023npjCM}], it is shown that the convergence with respect to the BZ sampling can be drastically improved, by combining a Monte Carlo integration with an appropriate interpolation of the screened potential between the calculated grid points, using the recently developed W-av method.~\cite{Guandalini2023npjCM} While W-av shares similarities with other methods,~\cite{Rasmussen_2016} it avoids sub-sampling of the BZ, gaining in computational efficiency.~\cite{Rasmussen_2016,daJornada_2017,Xia_2020}

In this paper, we present a general method  to efficiently combine the MPA and W-av schemes to compute QP band structures with coarse frequency and BZ samplings.
Overall, the MPA and W-av methods significantly reduce the computational cost of evaluating the $G_0W_0$ self-energy by acting  over the frequency and momentum integrals, respectively. 
%However, combining them is not straightforward.
%
Their combination, referred here as MPA@W-av, has been implemented in the \yambo{} package~\cite{yambo_2009,yambo_2019} and is applied to the calculation of the $G_0W_0$ band structure of graphene. 

The $GW$ quasi-particle band structure of graphene has been thoroughly studied, both in the neutral~\cite{Trevisanutto_2008,Yang2009} and doped regime.~\cite{Attaccalite_2009}
Here, the dominant effect of the $GW$ corrections to the density functional theory (DFT) band structure is an increase of the Fermi velocity, severely underestimated by DFT, and an enlargement of the energy separation between valence and conduction bands farther from the Dirac cone.
In the doped regime, $GW$ corrections are strongly reduced due to the efficient screening of the added electrons/holes.~\cite{Attaccalite_2009}
Close to the  Dirac (charge neutrality) point, the electron-electron interaction is expected to play an important role in intrinsic graphene, where the graphene Fermi velocity should present an ultraviolet logarithmic divergence due to the coexistence of Fermi points and strong correlation.~\cite{elias_dirac_2011,siegel2011many}
The divergence in the graphene self-energy is present in both Hartree-Fock and higher order theories, regardless of the screened or bare nature of Coulomb force.~\cite{siegel2011many,Barnes2014PRB}
A number of experimental~\cite{yu_interaction_2013,elias_dirac_2011,chae_renormalization_2012} and computational~\cite{trevisanutto2008ab,Attaccalite_2009} studies report the expected logarithmic renormalization of the graphene Fermi velocity, obtaining quantitative agreement with the theoretical results. However, due to the limitations regarding the momentum resolution, the purity of the samples, and residual doping, the scale at which the divergence sets in is difficult to assess.

On the {\it ab-initio} side, Ref.~[\onlinecite{trevisanutto2008ab}] reports $G_0W_0$ calculations with a 17\% increase of the Fermi velocity with respect to the DFT local-density approximation (LDA). Close to {the $K$ point}, the linear dispersion of the $\pi$ band is modified by the presence of a kink, attributed to low-energy $\pi \rightarrow \pi^*$ single-particle excitations and to the $\pi$ plasmon.
In Ref.~[\onlinecite{Attaccalite_2009}], a strong renormalization of the Fermi velocity at zero doping is reported, 
with a 50\% increase going from $0.846\times 10^6~m/s$ (LDA) to $1.25\times 10^6~m/s$ ($G_0W_0$).
Although the calculation methods are similar to the ones reported in Ref.~[\onlinecite{trevisanutto2008ab}], no kink is visible, and the bands recover the linear behavior at a $0.003$~\AA$^{-1}$ distance from $K$ (Ref.~\onlinecite{Attaccalite_2009}) instead of $0.7$~\AA$^{-1}$ (Ref.~\onlinecite{trevisanutto2008ab}), possibly as a result of a different BZ discretization. 

The $GW$ QP band structure of graphene has been also used as a starting point for the calculation of optical spectra within the GW+BSE method,~\cite{Hedin_1965,strinati1988application,Onida_2002} which revealed excitonic effects despite the semimetal screening properties of graphene.~\cite{Yang2009,Yang2011}
Indeed, graphene is a 2D material with a sharp dependence of the dynamical dielectric function on transferred momentum.
Moreover, the presence of two intense peaks (the $\pi$ and $\pi + \sigma $ plasmons) and an important low-energy contribution make it a prototypical case in which PPA may be inadequate to describe the screened interaction, as highlighted by electron-energy loss (EELS) experiments.~\cite{Kinyanjui2012,Wachsmuth2013,guandalini2023excitonic}
These properties make graphene an excellent test case for the proposed MPA@W-av method. 

In this work, as an application of the newly developed MPA@W-av scheme, we compute the $GW$ QP energies of graphene within the MPA@W-av approach and compare it 
with angle-resolved photoemission spectroscopy (ARPES) data.~\cite{Knox_2011} We also compare the efficiency of the MPA@W-av method with respect to the PPA and FF-RA computational schemes.
Moreover, the improved $\mathbf{k}$-resolution allows for an accurate calculation of the $GW$ Dirac cone, used here to determine the parameters of a generalized model of the logarithmic velocity renormalization. 

The work is organized as follows: 
In Section \ref{Sec_methods} the $GW$ methodology is introduced (Section~\ref{sec:gw}) and the MPA (Section~\ref{Sec_Mpa}) and W-av (Section~\ref{Sec_Wav}) approaches are presented. In Sec.~\ref{Sec_MPA_Wav} the combined MPA@W-av method is described 
and applied to obtain the QP band structure of graphene. 
The computational details are provided in Sec.~\ref{Sec_comp_details}.
In Secs.~\ref{Sec_dielectric} and \ref{sec:W(w)}, 
the dielectric properties of graphene are reported.
In Sec.~\ref{Sec_conv_k}, we address the convergence of the Fermi velocity and the gap at $M$  with a number of approximations and computational schemes. The frequency dependence of the graphene self-energy is studied in Sec.~\ref{Sec_self_energy}.
In Sec.~\ref{Sec_compare_exp}, we compare the obtained QP band structure of graphene with ARPES measurements.
In Sec.~\ref{Sec_dirac_cone} we investigate the renormalization of the Fermi velocity close to the Dirac point. In Sec.~\ref{Sec_conclusions} we draw the conclusions. Finally, in Appendix~\ref{Sec_Dirac_term} we present a static Dirac model of the long-wavelength limit of graphene irreducible polarizability,  while in Appendix~\ref{apendix_hf} we construct an hyperbolic Dirac model of the renormalization of the Fermi velocity.

\section{Methods and developments}
\label{Sec_methods}

\subsection{\texorpdfstring{The $G_0W_0$ self-energy}{The G0W0 self-energy} }
\label{sec:gw}

Within the framework of many-body perturbation theory, the quasi-particle energies are obtained by solving the QP equation:
\begin{equation}
\eig_{n\k}^{\mathrm{QP}} =  \eig_{n\k}+ \mel{n\k}{\SE(\eig_{n\k}^{\mathrm{QP}}) - v_{xc}}{n\k},
\label{Eq_QP_full}
\end{equation}
where $\eig_{n\k}$ and $|n\k\rangle$ are usually the KS energies and wavefunctions and  $v_{xc}$ is the exchange-correlation potential.
Alternatively, Eq.\eqref{Eq_QP_full} may be linearized by a Taylor expansion of the self-energy up to first order, leading to:
\begin{equation}\label{eq_QP_def}
\eig_{n\k}^{\mathrm{QP}} =  \varepsilon_{n\k}+ Z_{n \k}\mel{n\k}{\SE(\eig_{n\k}) - v_{xc}}{n\k},
\end{equation}
where the renormalization factor $Z_{n \bf k}$ is defined as:
\begin{equation}
   Z_{n\mathbf{k}} = \left[\left. 1-\langle n {\bf k}|\frac{\partial\Sigma(\omega)}{\partial \omega}| n {\bf k}  \rangle\right|_{\,\omega=\epsilon_{n {\bf k}}} \right]^{-1} .
   \label{Eq_QP_N}
\end{equation}

In order to obtain the QP correction of a single-particle state, $\ket{n\mathbf{k}}$, we need to compute its corresponding self-energy matrix element, $\SE_{n\k} \equiv \mel{n\k}{\SE}{n\k}$, in the $G_0W_0$ approximation.
Making explicit use of the Lehmann representation for $G_0$, in a Bloch plane-wave basis set, the diagonal matrix elements of the self-energy can be expressed as
\begin{multline}
   \SE_{n\k}(\o) = 
    -\sum\limits_m \sum\limits_{\G\G'}
    \int\frac{d\o'}{2 \pi i}  e^{i \o' \eta}
     \int\frac{d\q}{(2\pi)^3} \times \\[5pt]
     \frac{\dm_{nm}(\k,\q,\G)\scr^0_{\G\G'}(\q,\o')\dm^*_{nm}(\k,\q,\G')}
     {\o+\o'-\eig_{m\k-\q}+i\eta {\times} \sgn(\eig_{m\k-\q}-\mu)},
     \label{eq_GW_expl}
\end{multline}
where $\dm_{nm}(\mathbf{k},\mathbf{q},\mathbf{G})=  \mel{n\mathbf{k}}{e^{i(\mathbf{q}+\mathbf{G}) \cdot \mathbf{r}}}{m\mathbf{k-q}}$, and $\mu$ is the chemical potential.
$\scr^0_{\G\G'} = \inveps_{\G\G'}\vcoul_{G'}$ is the (non-self-consistent) screened interaction matrix in plane-wave representation, and $\inveps_{\G\G'}$ the inverse dielectric function.
The implicit limit $\eta \to 0^+$ ensures the proper time-ordering.
From now on, we drop the $0$ superscript of the screened interaction to ease the notation.

Notably, 
Eq.~\eqref{eq_GW_expl} contains a double integral in momentum and frequency space, which in principle requires the calculation of $W(\q,\o)$ over a sufficiently dense grid both in the BZ and in the frequency space.
In Eq.~\eqref{eq_GW_expl}
$W$ is the only operator whose analytical expression is not known and is typically split in two terms: the bare Coulomb interaction $v$, which is static, and the correlation part $W^c\equiv W-v$, which contains all the dynamical effects, i.e. the frequency dependence. This definition leads to a decomposition of the self-energy into exchange and correlation terms, $\SE_{n\k}(\o) \equiv \SEx_{n\k}+\SEc_{n\k}(\o)$.

%=============
\subsection{Frequency integration within MPA}\label{Sec_Mpa}
%=============

The correlation part of the screened Coulomb potential $W^c$  depends on frequency via the density-density
response function $\chi$, with $W^c(\omega)=v\chi(\omega) v$. In the MPA method, both $W^c$ and $\chi$ matrix elements are expressed as the sum of a small number of complex poles. For example, the components of $W^c$ are written as:
\begin{equation}
    W^{c}_{\G\G'}({\q},\omega) = \sum_{p=1}^{n_p} \frac{2 R_{p\G\G'} ({\q}) \Omega_{p\G\G'} ({\q})}{\omega^2-[\Omega_{p\G\G'}({\q})]^2},
    \label{eq:Xmp}
\end{equation}
where $\Omega_{p\G\G'}$ and $R_{p\G\G'}$ are the poles and residues, respectively, and $n_p$ is the total number of poles.
Since $v$ does not depend on the frequency, the poles of $W^c$ and $\chi$ are identical, and the residues of $W^c$ are the residues of the multipole model of $\chi$ scaled by the bare potential, $v$, on both sides.

As described in Ref.~[\onlinecite{Valido_2021}], for each momentum transfer $\bf q$ and each $\bf GG'$ matrix element,  poles and  residues are obtained through a non-linear interpolation with frequency points sampled in the complex plane, $z\equiv \omega+i\varpi$:

\begin{equation}
    \sum\limits_{p=1}^{n_p}
    \frac{2\Rp_{p\G\G'}(\q)\Op_{p\G\G'}(\q)}{z_i^2-[\Op_{p\G\G'}(\q)]^2} = \scr^{\text{c}}_ {\G\G'}(\q,z_i), \ 
    \label{Eq_MPA_av}
\end{equation}
where $i=1,\ldots,2n_p$ and the set of complex frequencies $z_i$ is conveniently selected according to the doubled parallel sampling defined in Ref.~[\onlinecite{Valido_2021}].
The correlation part of the $GW$ self-energy is then integrated analytically in the frequency space, leading to:
\begin{multline}
\SE^{c}_{n\k}(\o) = 
    -\sum\limits_m \sum\limits_{\G\G'}\sum\limits_{p=1}^{n_p}
     \int\frac{d\q}{(2\pi)^3} \times \\
     \frac{\dm_{nm}(\k,\q,\G)R_{p\G\G'}(\q)\dm^*_{nm}(\k,\q,\G')}
     {\o+\Omega_{p\G\G'}({\q})-\eig_{m\k-\q}+i\eta {\times} \sgn(\eig_{m\k-\q}-\mu)}.
    \label{eq:SEmp}
\end{multline}
This expression generalizes the PPA solution to the case of a multipole expansion for $\scr^\text{c}$. It bridges between PPA, in the case of one pole, and an exact full-frequency approach, for an increasing number of poles. More details about this procedure can be found in Ref.~[\onlinecite{Valido_2021}].

%==============
\subsection{Momentum integration within W-av}\label{Sec_Wav}
%==============
The W-av approach extends to the dynamically screened potential $W$ the Monte Carlo integration
used to treat the bare Coulomb interaction~\cite{yambo_2009,Guandalini2023npjCM} when integrated over transferred momentum. 
In Ref.~[\onlinecite{Guandalini2023npjCM}] the method was developed for the case of 2D semiconductors treated at the  PPA level (equivalent to Eq.~\eqref{eq:SEmp} with $n_p=1$ and real poles), here denoted as PPA@W-av.
The self-energy integral was discretized over the BZ by exploiting the smooth dispersion of $\dm_{nm}(\k,\q,\G)$ and $\Omega_{\G\G'}^{\text{PPA}}({\q})$, with respect to ${\q}$. However, the residues $R_{\G\G'}^{\text{PPA}}(\q)$ of the screened potential vary rapidly due to the divergence of the Coulomb interaction.

To account for this, the integrand in Eq.~\eqref{eq:SEmp} is split into slowly and rapidly varying components with respect to $\q$.
The rapidly varying components, proportional to the correlation part of the static screened potential $W^c$, are first interpolated in order to determine $\scrc$ between the grid points and then used in a Monte Carlo integration, to compute the average value of $\scrc$ in the area around each grid point. The computed average, $\overline{\scrc}$, is finally included in the integral, providing a significant computational speedup.~\cite{Guandalini2023npjCM}
In Ref.~[\onlinecite{Guandalini2023npjCM}] it was shown that within PPA, the use of this averaged screened potential in the static limit, $\omega=0$, significantly accelerates the convergence of the self-energy integral over the momentum transfer with respect to the number of $\k$-points in the discretization of the BZ.

%==============
\subsection{Combining the MPA and W-av methods}\label{Sec_MPA_Wav}
%==============
%
The key aspect of MPA is the use of optimal sampling strategies for $\chi$ in the frequency complex plane, which allows one to reach full-frequency accuracy by sampling only a few frequency points. Conversely, the W-av method allows one to obtain accurate quasi-particles for 2D materials with coarse ${\bf k}$-point grids. Combining the two methods  allows for convergence acceleration in both frequency and momentum space.
However, some caution must be taken, since the hypotheses used in PPA@W-av about the $\q$ dependence of the position and the amplitude of the plasmon poles may not apply to the MPA description.

While W-av interpolates a monotonic function in a small neighbourhood of each point in the BZ, MPA interpolates a complex structure in the frequency space. This suggests that the MPA interpolation is sufficiently flexible to model the frequency dependence of $\overline{\scrc}(\o)$ resulting from the W-av procedure. 
For this reason, we invert, in the case of MPA@W-av, the order of the self-energy integrals over frequency and momentum with respect to the PPA@W-av approach described in Ref.~[\onlinecite{Guandalini2023npjCM}].
We begin by using W-av to compute the average screened interaction $\overline{\scrc}(\o)$ in the discretized BZ for all frequency points of the MPA sampling. 
Next, we interpolate the results in frequency space using a multipole model.
Finally, we evaluate the frequency integral in the self-energy analytically using MPA. The step-by-step procedure is outlined in detail in the following paragraphs.

Let's consider separately the exchange and correlation terms of the self-energy, $\SEx_{n\k}$ and $\SEc_{n\k}(\o)$.
Since $\vcoul$ is frequency independent, $\SEx_{n\k}$ can be easily integrated analytically in the frequency space, while the integral over momentum transfer can be computed on a Monkhorst-Pack grid~\cite{Monkhorst_1976} by means of the v-av method:~\cite{yambo_2009,Guandalini2023npjCM}
\begin{equation}
\SEx_{n\k} = -\frac{1}{N_{q}\Omega}\sum\limits_{v}^{\text{occ}}
\sum\limits_{\q,\G} |\dm_{nv}(\k,\q,\G)|^2 \bar{\vcoul}_{\G}(\q) \ ,
\end{equation}
where $\Omega$ is the unit cell volume and $\overline{\vcoul}$ is the average Coulomb interaction within a region of the BZ centred around $\q$ (here reported for a 2D case), 
\begin{equation}
    \overline{\vcoul}_\G(\q) =
    \frac{1}{D_{\q}}\int\limits_{D_{\q}}
    \frac{d\q'}{(2\pi)^2}
    {\vcoul}_\G(\q'),
    \label{Eq_v_av}
\end{equation}
where $D_{\q}$ is the integration domain centred around $\q$ within the 2D Monkhorst-Pack grid.

For the correlation part of the self-energy, $\SEc_{n\k}(\o)$, we split the integrand over $\q$ following a procedure similar to the PPA@W-av in Ref.~[\onlinecite{Guandalini2023npjCM}]. 
First, in analogy with Eq.~\eqref{Eq_v_av}, we define $\overline{W^c}_{\G\G'}$ as the average screened Coulomb interaction, which is the term varying rapidly wrt $\mathbf{q}$:
\begin{equation}
    \overline{W^c}_{\G\G'}(\q,\o) =
    \frac{1}{D_{\q}}\int\limits_{D_{\q}}
    \frac{d\q'}{(2\pi)^2}
    {W^c}_{\G\G'}(\q',\o).
    \label{Eq_W_av}
\end{equation}
This allows us to write the self-energy as:
\begin{multline}
    \SEc_{n\k}(\o) = 
    \frac{-1}{N_q\Omega}
    \sum\limits_m \sum\limits_{\G\G'\q}
  \int\frac{d\o'}{2 \pi i}e^{i \o' \eta}
    \times \\
     I_{\G\G'}^m(\q,\o+\o')
     \overline{\scrc}_{\G\G'}(\q,\o').
\end{multline}
where $I_{\G\G'}$ is a term that varies slowly with $\q${:}
\begin{equation}
    I_{\G\G'}^m(\q,\o+\o') {\equiv}
    \frac{\dm_{nm}(\k,\q,\G)\dm^*_{nm}(\k,\q,\G')}
     {\o+\o'-\eig_{m\k-\q}+i\eta {\times} \sgn(\eig_{m\k-\q}-\mu)}
\end{equation}
Next, we build a multipole representation analogous to Eq.~\eqref{eq:Xmp} but for the average potential $\overline{W^c}(\o)$.
The nonlinear interpolation is then solved with the method developed in Ref.~\onlinecite{Valido_2021} based on Pad\'e approximants.
After integrating in both momentum and frequency, we obtain:
\begin{widetext}
\begin{equation}
    \SE^{c}_{n\k}(\o) =
    \frac{1}{N_q\Omega}\sum\limits_m\sum\limits_{\G\G'\q}\sum_{p=1}^{n_p} \, %\\
    \frac{\dm_{nm}(\k,\q,\G) \overline{R}_{p\G\G'}(\q)\dm^*_{nm}(\k,\q,\G')}{\o-\eig_{m\k-\q}+(\overline{\Op}_{p\G\G'}(\q)-i\eta)\times \sgn(\mu-\eig_{m\k-\q})},
    \label{Eq_SE_MPA_Wav}
\end{equation}
\end{widetext}
where $\overline{\Omega}_{p}$ and $\overline{R}_{p}$ are the poles and the residues of the multipole model for the average potential, $\overline{\scrc}$.

Equation~\eqref{Eq_SE_MPA_Wav} allows us to simultaneously address the convergence problems with respect to the BZ sampling and the complex frequency dependence of the correlation part of the screened interaction $\scrc(\q,\o)$, as demonstrated in the next sections for the case of graphene.
We emphasize that, although in this paper we devise the formulation for 2D semiconductors and semimetals, the proposed method can be used to treat systems with different dimensionalities (1D and 3D) and can be extended to systems with different screening properties (e.g. metals).

%%%%%%%%%%%%%%%%%%%%%%%%%%
\section{Computational details}\label{Sec_comp_details}
%%%%%%%%%%%%%%%%%%%%%%%%%%

DFT calculations were performed using the plane wave  implementation of the Quantum ESPRESSO package,~\citep{QE_2020} within the local-density approximation (LDA) exchange-correlation functional.~\citep{LDA} 
We adopted norm-conserving pseudopotentials to model the electron–ion interaction and the kinetic energy cutoff for
the wavefunctions was set to $90$ Ry.
$G_0W_0$ calculations were performed with the \yambo{}  package.~\citep{yambo_2009,yambo_2019}
The QP energies have been calculated with the linearized expression in Eq.~\eqref{Eq_QP_N}.
The derivative of the self-energy, required to compute the $Z_{n\k}$ factors, has been computed by finite differences using a frequency interval of $\Delta \o = 0.1$ eV. 
In the calculation of the Green's function,
we used a finite damping $\eta = 0.01$~eV.
For all calculations other than the QP band structures in Figs.~\ref{Fig_exp}-\ref{fig:PPA-MPA-model}, we used a cutoff of $5$ Ry for the size of the dielectric matrix, including up to $100$ states in the sum-over-state of the response function.
The same number of states has been employed in the calculation of the correlation part of the self energy.

In the calculation of the QP band structures of Figs.~\ref{Fig_exp}-\ref{fig:PPA-MPA-model}, we used a cutoff of $10$ Ry with $200$ states both in the calculation of the dielectric matrix and  the self-energy.
When working with the PPA model, we adopted the Godby-Needs scheme~\cite{Godby_1989} by sampling the polarizability in the complex frequency space at $z_1 = 0$ and $z_2 = i$ Ha.
In the MPA implementation, we used a double parallel sampling with shifts $\varpi_1 = 0.02$ Ha and $\varpi_2 = 1$ Ha (see Refs.~[\onlinecite{Valido_2021,leon2023efficient}] for further details), with $10$ poles and a quadratic sampling along the real frequency axis, as explained in Ref.~[\onlinecite{leon2023efficient}].
When not explicitly declared, we used a $60\times 60$ Monkhorst-Pack grid to sample the BZ.
The Coulomb interaction has been truncated with a slab cutoff~\cite{Beigi_2006, Rozzi_2006} in order to remove interactions within supercell replicas both in $G_0W_0$ calculations. 
This allowed us to use a supercell with an interlayer distance of $L = 10~$\AA{} to obtain converged results with respect to the cell size.
%\AG{Otherwise explicitly defined, we used a smearing of...}

The calculations to address the Fermi velocity renormalization in Sec.~\ref{Sec_dirac_cone} were performed with the method described in Ref.~[\onlinecite{Attaccalite_2009}], which allows us to include extra $\mathbf k$-points to the regular grids used for the polarizability. 
%\DVcancel{where the quasi-particle energies are evaluated}. 
For such calculations we made use of a finite occupation smearing consistent with the adopted $\mathbf{k}$ grid, as discussed in Sec.~\ref{Sec_dirac_cone} and detailed in the \suppinfo. 
%\DVnote{forse quest'ultima frase non è chiarissima. Inoltre bisogna rimandare alle supp.info per il ruolo dello smearing}

%%%%%%%%%%%%%%%%%%%%%%%%%%
\section{Results and discussion}\label{Sec_results}
%%%%%%%%%%%%%%%%%%%%%%%%%%

We now present the results obtained with the MPA@W-av scheme derived in Sec.~\ref{Sec_methods} for the case of graphene.
In particular, we compute the QP energies and examine the frequency dependence of the response functions and the self-energy of graphene. 
Results computed at the MPA@W-av level are then compared with those obtained from PPA and FF-RA, and the computed $GW$ band structures compared with angle resolved photoemission spectroscopy (ARPES) measurements.~\cite{Knox_2011}
Considering the semimetallic character of graphene, in App.~\ref{Sec_Dirac_term} we also present a scheme based on a static Dirac model to include the long-wavelength limit contribution of the Dirac cone into the irreducible polarizability, a term which is often missing in common $GW$ implementations.
The effect of this scheme is also discussed.

%=====================================
\subsection{Dielectric properties of graphene}\label{Sec_dielectric}
%=====================================

\begin{figure}%[hbtp]
\includegraphics[width=0.48 \textwidth]{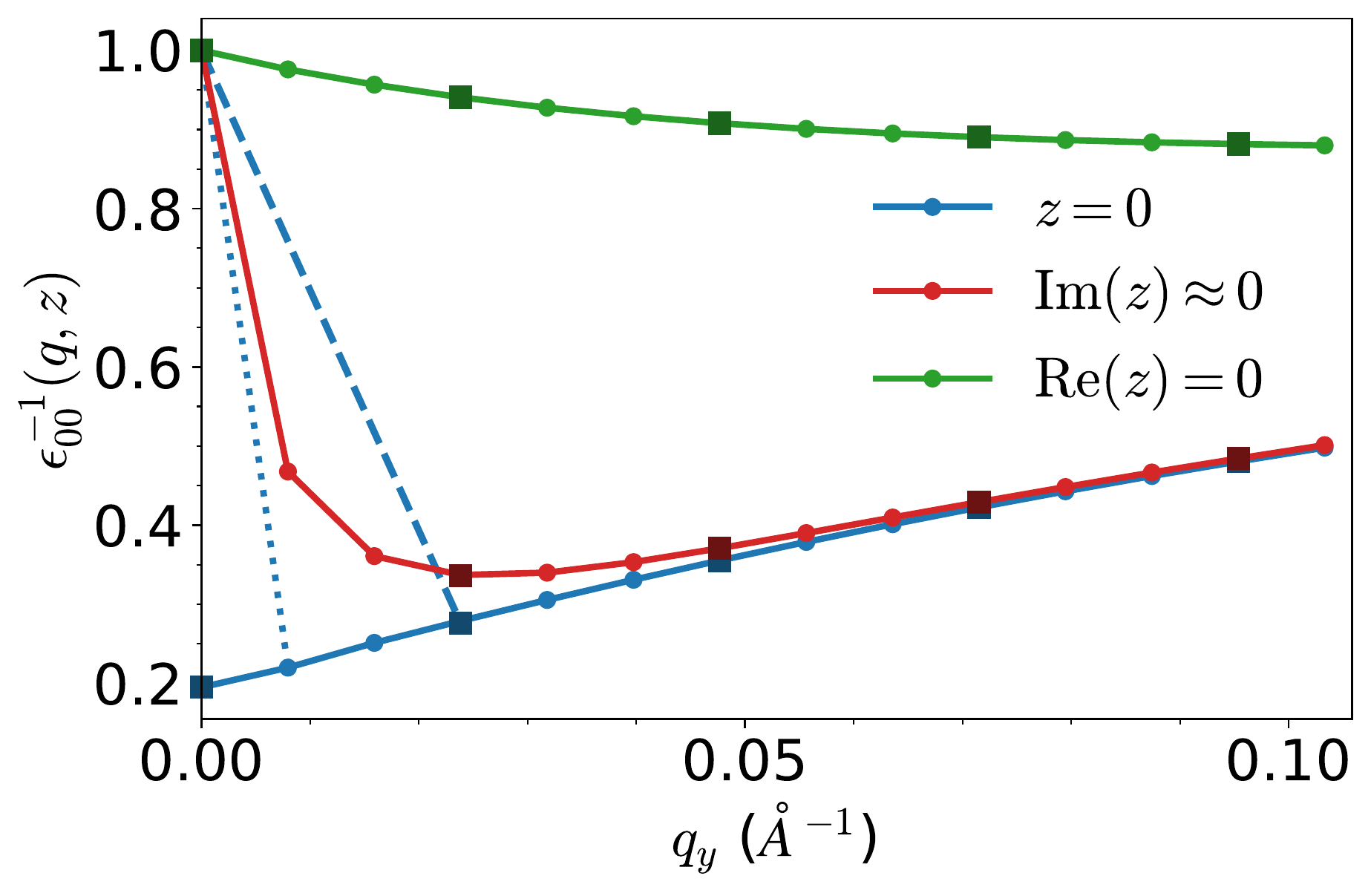}
\caption{(Color online) Momentum dependence of the real part of the inverse dielectric function of graphene along the $\Gamma M$ direction, calculated for $\G=\G'=0$ for three different frequencies in the complex plane $z = \o + i\varpi$: in blue the static limit ($\o = \varpi  = 0.$ eV), in red near the real axis ($\omega \approx 0.16$ eV, $\varpi \approx 0.54$ eV) and in green on the imaginary axis ($\omega = 0$ eV, $ \varpi \approx 27$ eV).
Dark squares (dots) indicate values obtained with a $60\times 60\times 1$ ($180\times 180 \times 1$) grids.
The dashed lines are the static inverse dielectric function calculated not including the intra-band contribution in the $\q\to 0$ limit. 
\label{Fig_epsm1}}
\end{figure}

In this Section we discuss some key features of the screening of graphene that are helpful to understand the subsequent results.
In Fig.~\ref{Fig_epsm1} we plot the macroscopic component ($\G=\G'=0$) of the inverse dielectric matrix of graphene as a function of momentum transfer, calculated at three different points in the complex frequency plane, $z \equiv \o + i\varpi$. The symbols represent calculations that include the long-wavelength contribution of the Dirac cone, derived in Sec.~\ref{Sec_Dirac_term}, while the dashed/dotted lines indicate calculations without this contribution.

In the static case (blue symbols), the inverse dielectric function approaches the value $\inveps \approx 0.2$ linearly in the long-wavelength limit.
This is consistent with the Dirac model,~\cite{Shung_1986,Gorbar_2002,Ando_2006,Wunsch_2006,Barlas_2007,Wang_2007,Sarma_2007} that is a good approximation in this regime.
When the  $\q\to 0$ limit contribution is not explicitly included (dashed blue line) the screening function of graphene behaves as a 2D semiconductor, recovering the correct semi-metallic limit only for infinite $\q$-point grids. It is evident that such a behaviour drastically slows down the convergence of the QP corrections.

At finite frequencies but close to the real axis (red symbols), the inverse dielectric function approaches one very quickly in the long wavelength limit, as expected for a 2D system.
In Fig.~\ref{Fig_epsm1} we present values for $\inveps$ obtained with two different  Monkhorst-Pack grids. The comparison between the results obtained with the $60\times 60$ Monkhorst-Pack grid (squares) and the $180\times 180$ grid (circles) shows that very dense grids are needed to describe the sharp behaviour for small  ${\bf q}$.
The inverse dielectric function calculated at an imaginary frequency (green symbols) shows instead a smooth behaviour with respect to momentum transfer, since $\epsilon^{-1}$ is sampled far from its poles, that lie on the real frequency axis.

The PPA and MPA methods resort to different frequency samplings. PPA samples the frequency plane only in the static limit and on the imaginary axis (illustrated here by the smooth blue and green curves), while MPA also samples the polarizability at finite frequencies near the real axis (rapidly varying red curve). Since W-av is designed to improve the description of rapid variations with respect $\q$ using coarse grids,~\cite{Guandalini2023npjCM} we expect the W-av method to be more relevant for MPA than for PPA.

\begin{figure}%[hbtp]
\includegraphics[width=0.48 \textwidth]{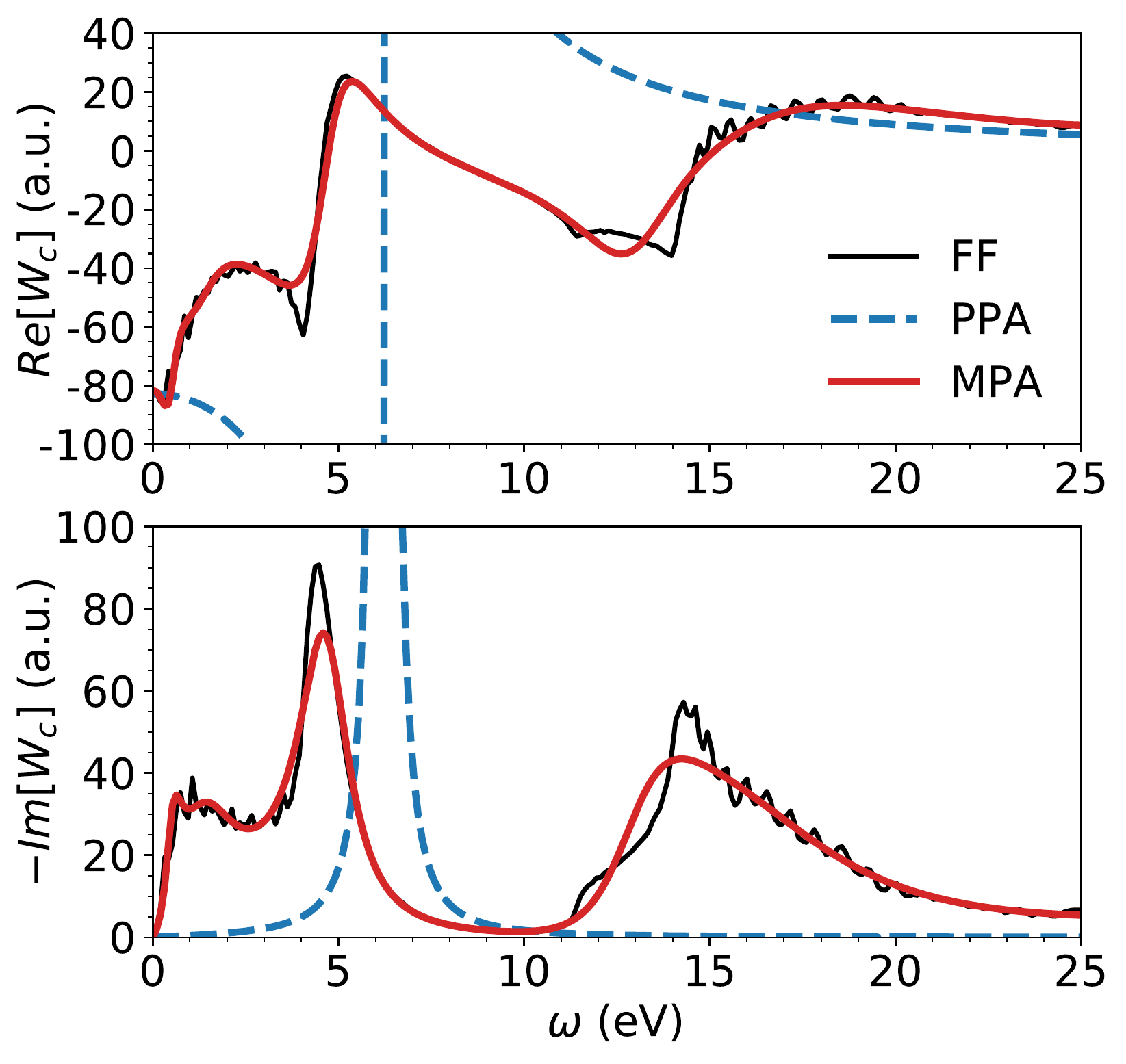}
\caption{Frequency dependence of the real (upper panel) and imaginary (lower panel) part of the correlation screened interaction ($\G=\G'=0$) of graphene computed along a line parallel to the real frequency axis, at a distance of $\varpi = 0.1$ eV. The momentum transfer is %\DALchange{$q \approx 0.05$ a.u.}
{$q \simeq 0.085$~\AA$^{-1}$} along the $\Gamma M$ direction. 
Full frequency grid (FF), plasmon-pole approximation (PPA) and  multipole approximation (MPA) are represented with black, blue and red lines respectively.}
\label{Fig_W}
\end{figure}

%%===============
\subsection{Frequency dependence of the screened potential}
\label{sec:W(w)}
%%===============

In Fig.~\ref{Fig_W}, we show the frequency dependence of the macroscopic matrix element $\G=\G'=0$ of the screened interaction $\scrc$ computed at finite momentum transfer 
%\DALchange{$q \approx 0.05$ a.u.}
$q \simeq 0.085$~\AA$^{-1}$ along the $\Gamma M$ direction within the random phase approximation (RPA). %level.
In the plot, we compare the PPA (dashed blue line) and MPA (red line) description against a direct evaluation on the frequency grid  (black line). $W^c$ was evaluated at frequencies $\omega+i\varpi$ parallel to the  real frequency axis, with $\varpi = 0.1$ eV.
The imaginary part of the macroscopic screened interaction is proportional to the electron energy-loss spectrum, already discussed in the literature.~\cite{Despoja_2012,Despoja_2013,Novko_2015,Nazarov_2015,Li_2017,guandalini2023excitonic}
In agreement with previous results, $-\text{Im}[\scr]$ presents three main features: a low-energy shoulder, the $\pi$ plasmon (around $5$ eV) and the $\pi+\sigma$ plasmon (around $15$ eV).

The PPA approximation captures the static and high-frequency limits of the real part of the screened interaction. However, it fails to capture the qualitative behaviour of the screened interaction, with its pole located between the $\pi$ and $\pi+\sigma$ plasmons, around $\Omega_{\text{PPA}} \approx 6.22$~eV.
It is known that PPA fails to describe QP properties~\cite{Giantomassi2011PSSB,Miglio2012EPJB,Valido_2021} in systems presenting structured features of $W$ at the low-energies, such as metals.~\cite{leon2023efficient} 
That is also the case of graphene.
In contrast, the MPA approximation, here evaluated considering ten poles, successfully reproduces all the key features of both the real and imaginary parts of the screened interaction. It is important to note that the computational cost of MPA can be effectively compared to that of conventional FF treatments by evaluating the number of frequency sampling points to be used. Specifically, MPA requires twice the number of poles (i.e., order of tens) while the FF integration on the real axis requires hundreds or thousands of them, depending on the frequency structure of the screened potential.~\cite{Valido_2021}

\begin{figure}
\includegraphics[width= 0.48\textwidth]{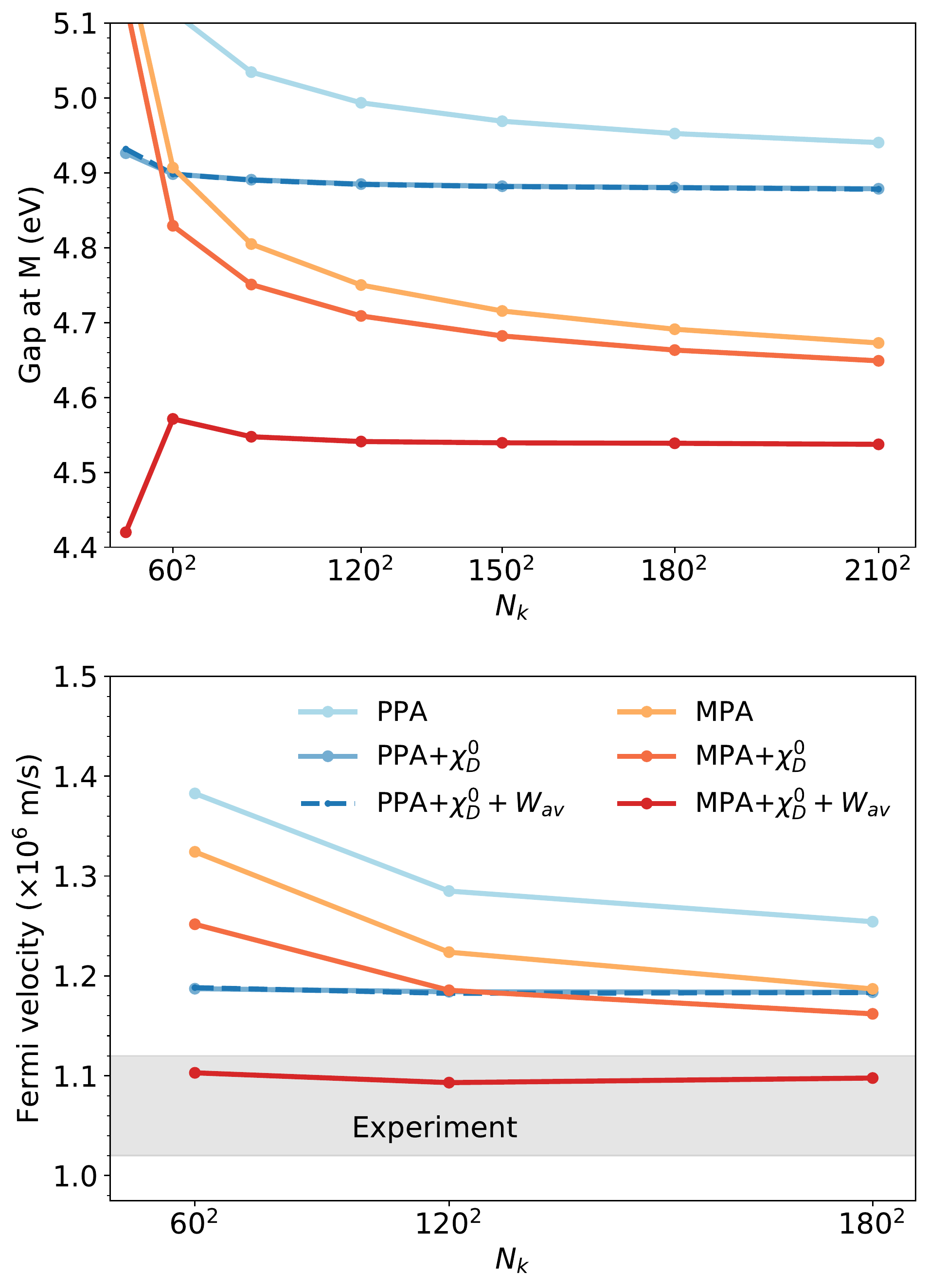}
\caption{Convergence of the quasi-particle band gap at $M$ (top) and Fermi velocity (bottom) of graphene with respect to the number of sampling points in the BZ.
The blue lines indicate results obtained with the plasmon-pole approximation (PPA).  
The dark (light) blue line refers to calculations done with (without) the inclusion of the long-wavelength contribution $\chi^0_D$ to the irreducible polarizability, according to Eq.~\eqref{Eq_chi_0_D_stat}. The dark dashed blue line indicates results obtained considering the inclusion of  $\chi^0_D$ and the use of the W-av method.
Yellow, orange and red lines indicate results obtained with the multipole approximation (MPA).
Orange lines indicate results where the long-wavelength contribution $\chi^0_D$ is added to the irreducible polarizability, while red lines indicate results obtained considering the inclusion of $\chi^0_D$ and the use of the W-av method. 
}
\label{Fig_gap}
\end{figure}

%============================
\subsection{Convergence of W-av with the k-mesh}\label{Sec_conv_k}
%============================

\begin{figure}%[hbtp]
\includegraphics[width=0.48 \textwidth]{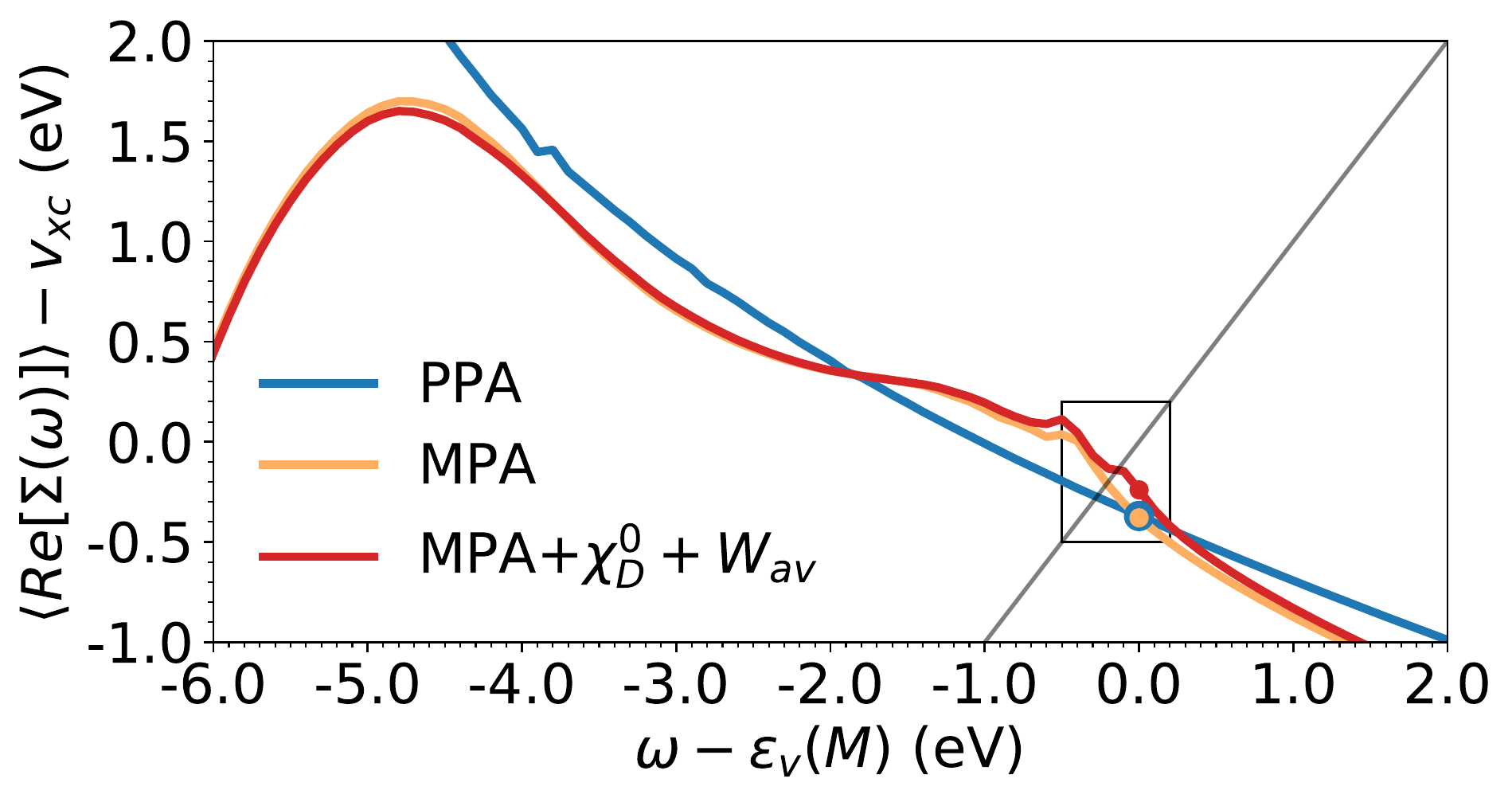}
\caption{Frequency dependence of the real part of the self-energy of graphene, calculated at the top of the valence band at the $M$ point.
The blue line is obtained with the plasmon-pole approximation (PPA), the yellow line with the multipole approximation (MPA), the red line indicates the MPA result when $\chi^0_D$ is included and the W-av method is applied.
Coloured dots indicate the self-energy computed at the KS energy, i.e. $\o'= \o-\eig_v(M) = 0$.
The grey straight line represents the condition $\Sigma(\o)-v_{\mathrm{xc}} = \o-\eig_v(M)$, thus providing the graphical solution of Eq.~\eqref{Eq_QP_full}.
}
\label{Fig_SE}
\end{figure}

The number of $\k$-points ($N_k$) used for discretizing the BZ is a challenging convergence parameter for 2D semiconductors, due to the sharp behaviour of the inverse dielectric function as a function of momentum transfer.~\cite{qiu2016screening,Huser_2013}  As shown in Sec.~\ref{Sec_dielectric} this is also the case of graphene.  
In the top panel of Fig.~\ref{Fig_gap}, we show the results of the QP gap at the $M$ point computed with different approximations.
Both the PPA and MPA methods (without the inclusion of the intraband correction $\chi^0_D$ (see App.~\ref{Sec_Dirac_term}), nor the use of the W-av method) show a very slow convergence with respect to $N_k$. This is due to the poor description of the long-wavelength limit of the static inverse dielectric function, as explained in Sec.~\ref{Sec_dielectric}.
The inclusion of $\chi^0_D$ greatly accelerates the convergence of the PPA gap, but the convergence of the MPA gap remains slow, showing that the Dirac cone contribution has a different impact on the two approximations. 

In PPA, the $\chi^0_D$ contribution correctly smooths out the static limit and the screened interaction is evaluated only at $\o =\varpi=0$ and $\o = 0$ \& $\varpi \approx 27$~eV, where the inverse dielectric matrix is smooth with respect to momentum transfer.
Conversely, in MPA, the screened interaction is also evaluated at frequencies close to the real axis, where the inverse dielectric function exhibits a rapid variation in the long-wavelength limit (see Fig.~\ref{Fig_epsm1}).
Only when MPA is applied in combination with W-av, the convergence of the QP properties is greatly accelerated. In fact, for a grid of $60\times 60$ $\mathbf{k}$-points, the difference in the QP gap with respect to the converged value is around $30$ meV, while without W-av the error is larger than $300$ meV.
In the bottom panel of Fig.~\ref{Fig_gap}, we show the convergence of the Fermi velocity, evaluated at the $K$ point through a first-order finite-difference formula   
%\DALcancel{along the $\Gamma K$ direction} 
with $\Delta k = 0.085$~\AA$^{-1}$, chosen to be 
%along the $\Gamma K$ direction,
consistent with determination of the Fermi velocity done from experiments.~\cite{Knox_2011}
The considerations  previously done for the gap at $M$ hold also for the Fermi velocity: PPA and MPA results converge slowly with respect to $\mathbf{k}$-points and 
the inclusion of $\chi_D^0$ greatly accelerates the convergence for PPA, without a significant effect for MPA.
Again, MPA@W-av with $\chi_D^0$ greatly accelerates the convergence.

\begin{figure}%[hbtp]
\includegraphics[width=0.48 \textwidth]{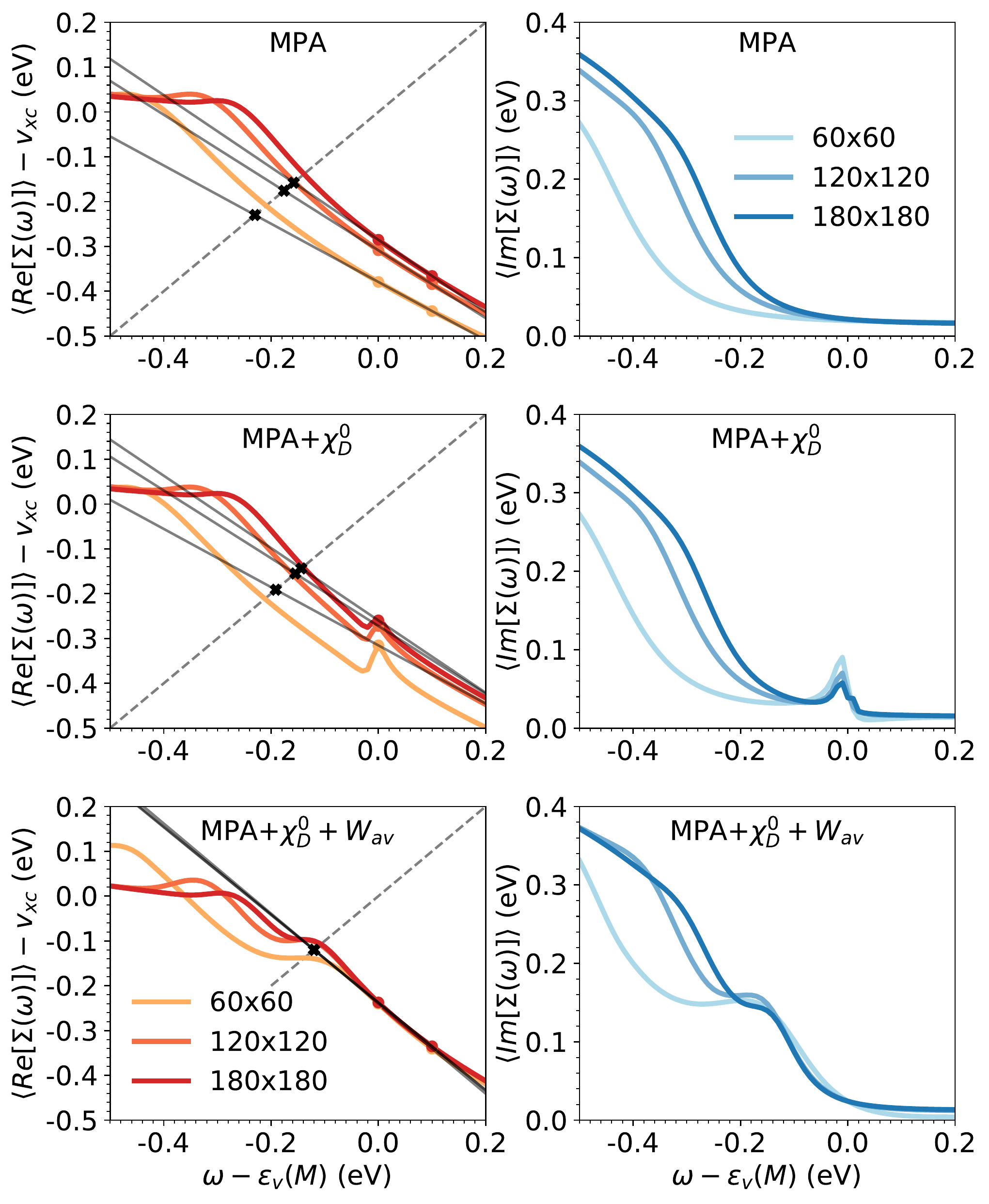}
\caption{Zoom of the frequency dependence of the graphene real (left) and imaginary (right) part of the self-energy  calculated for the top of the valence band at the $M$ point (black box in Fig.~\ref{Fig_SE}). From the top to the bottom: results obtained with the multipole approximation (MPA), MPA plus the long-wavelength contribution $\chi^0_D$ (see Eq.~\eqref{Eq_chi_0_D_stat}), MPA including the $\chi^0_D$  contribution and application of the W-av method. Yellow (light blue) lines indicate results obtained with the $60\times 60$ grid, orange (blue) lines  with the $120\times 120$ grid, while red (dark blue) lines obtained with the $180\times 180$ grid.
The grey shaded lines match the condition $\Sigma(\o)-v_{\mathrm{xc}} = \o-\eig_v(M)$, thus represents the graphical solution of Eq.~\eqref{Eq_QP_full}.
The grey lines represent the linearization of the self-energy of Eq.~\eqref{Eq_QP_N}.
Black crosses represent the linearized solution of the QP equation [see Eq.~\eqref{Eq_QP_N}].
}
\label{Fig_SE_zoom}
\end{figure}

%======================
\subsection{The self-energy of graphene}\label{Sec_self_energy}
%======================
. 
Given a self-energy, the reliability of the quasi-particle correction and lifetime of the state $|n\k\rangle$ depends on the accuracy of the real and imaginary parts of the self-energy at energies close to the corresponding Kohn-Sham eigenvalue $\eig_n(\k)$, as indicated by either Eq.~\eqref{Eq_QP_full} or Eq.~\eqref{Eq_QP_N}.
In Fig.~\ref{Fig_SE}, we show the frequency dependence of the graphene self-energy calculated at the top of the valence band at the $M$ point, within PPA and MPA. The effect of W-av used with MPA is also shown. For the sake of clarity, the frequency axis is shifted with respect to the KS eigenvalue corresponding to the top of the valence band at M, i.e. $\o' = \o-\eig_v (M)$.

As discussed in Section~\ref{sec:W(w)}, within PPA the plasmon energy is evaluated to be around $6.22$~eV, thus, only a region of the tail of the PPA self-energy is represented within the range of the plot in Fig.~\ref{Fig_SE}.
In contrast, the MPA self energy shows a complex structure in the same energy range, both around the QP solution and  the energy region associated with the $\pi$ plasmon ($\o' \sim -5$~eV).
The value of the self-energy obtained with PPA and MPA at $\omega = \eig_v(M)$ is very similar, while its derivative differs due to the different behaviour of $\Sigma(\o')$ at small negative frequencies.
Therefore, the difference between the PPA and MPA solutions of the linearized QP equation, Eq.~\eqref{eq_QP_def}, is mainly due to the $Z$ factors.
The use of $\chi^0_D$ with W-av impacts the MPA results through the low energy features of the self-energy, corresponding to the poles of $W$ at low energies and transferred momenta, where the inverse dielectric function varies quickly with $\q$ (see Fig.~\ref{Fig_epsm1}). In particular, $\chi^0_D$ adds an extra peak around $\o'=0$, related to the vanishing energy transition close to the Dirac point.

In Fig.~\ref{Fig_SE_zoom}, we present a zoom of the frequency dependence of the graphene self-energy, real (left panel) and imaginary (right panel) components. The self-energy is calculated within an energy range close to the solution of the QP equation, using MPA for different $\k$-meshes with and without $\chi^0_D$ and W-av.
Without the inclusion of the $\chi^0_D$ term (top panels), graphene behaves as a 2D semiconductor with decreasing gap as the grid size increases. 
The onset in the imaginary part of the self-energy approaches zero for denser grids.
However, the convergence with respect to the grid size is very slow, since the spurious gap $\Delta = 2\gamma\Delta q$ closes with the inverse of the number of sampling point in each periodic direction (e.g. $n_q$ for a $n_q\times n_q$ grid), similarly to the case of metals.~\cite{Cazzaniga_2008}

Likewise, the value of the real part of the self-energy, $\text{Re}[\SE_{n\k}(\o'=0)]$, and its derivative, $Z_{n\k}$,  also converge slowly, as shown by the solid black lines in the top left panel.
The pole, added when including $\chi^0_D$ (central panels), remains in the same energy position and decreases in intensity for larger grids.
This pole is located precisely in the region where $Z_{n\k}$
is evaluated, making it numerically unstable. As shown in Fig.~\ref{Fig_SE_zoom}, the calculation of $Z_{n\k}$ by a Newton difference quotient significantly depends on the frequency interval used in the numerical differentiation. For this reason, to compute the QP energy, we first calculate $Z_{n\k}$  disregarding the Dirac cone contribution and only afterwards include  $\chi^0_D$ to evaluate $\SE_{n\k}(\o'=0)$.

Finally, when both $\chi^0_D$ and  W-av are included (bottom panels), the pole introduced by  $\chi^0_D$ is partially overlapped with the onset in the imaginary part of the self-energy, resulting in a smother frequency-dependence. The smoothing of the onset increases for denser $\k$-grids, since the W-av method averages the transition energies in a small domain around each $\q$.
This averaging stabilizes the numerical evaluation of the $Z_{n\k}$ factor in the $\q \to 0$ limit and is able to accelerate the convergence of the solution of the QP equation with respect to the BZ sampling.
Since $\chi^0_D$ is static, it is its combination with the W-av approach that provides an effective dynamical correction to the MPA self-energy, improving the accuracy of the QP energy. However, if we want to go beyond the determination of the QP energies and improve on the description of the imaginary part of the self-energy in the region corresponding to $\omega \lesssim -0.2$~eV, we would need either a denser $\k$-grid or an accurate dynamical treatment of the long-wavelength limit, e.g., through a dynamical extension of the intraband $\chi^0_D$ model.

Last, we discuss the computational advantage in using the MPA@W-av approach.
Despite the fact that we have not explicitly performed converged calculations with FF in the real axis formulation, we may reasonably assume that around 1000 frequency steps would have been required ($N_{\omega} = \omega_{\mathrm{max}}/\Delta \omega \approx 4 \text{Ha}/0.1 \text{eV} \approx 1000$).
The computational gain due to MPA is thus approximately a factor $50$.
Using a $60\times 60$ grid already provides converged results with W-av, while the $210 \times 210$ grid is still too coarse to obtain converged gaps without. Taking the $\mathbf{k}$ and $\mathbf{q}$ grids to be identical, $G_0W_0$ typically scales as $N^2_k$. When this is the case, within W-av the computational cost of a $G_0W_0$ calculation is reduced by at least a factor $150$.
Thus, the proposed combined method (MPA@W-av) has a total computational speedup of $50 \times 150 = 7500$, with respect to more straight forward approaches.

\begin{table}
\centering
\begin{ruledtabular}
\begin{tabular}{ccccc}
\\[-3pt]
& LDA & PPA & MPA & Exp. \\[5pt]
\hline\\[-3pt]
$\gamma$ ($10^6$ m/s) & $0.88$ & $1.18$ & $1.09$ & $1.07 \pm 0.05$ \\
Gap $M$ (eV)  & $4.00$ & $4.88$ & $4.54$  \\
\end{tabular}
\end{ruledtabular}
\caption{Fermi velocity $\gamma$ and gap at $M$ of graphene calculated with $DFT$ and $G_0W_0$ with the plasmon-pole approximation (PPA) and 
the multipole approximation (MPA).
The experimental measurement of the Fermi velocity from Ref.~\onlinecite{Knox_2011} is also reported. Other results from the literature are mentioned in the main text.
}
\label{Tab_QP}
\end{table}

%%%%%%%%%%%%%%%%%%%%%%%%%%%%%%%%%%

\begin{figure}%[hbtp]
\includegraphics[width=0.48 \textwidth]{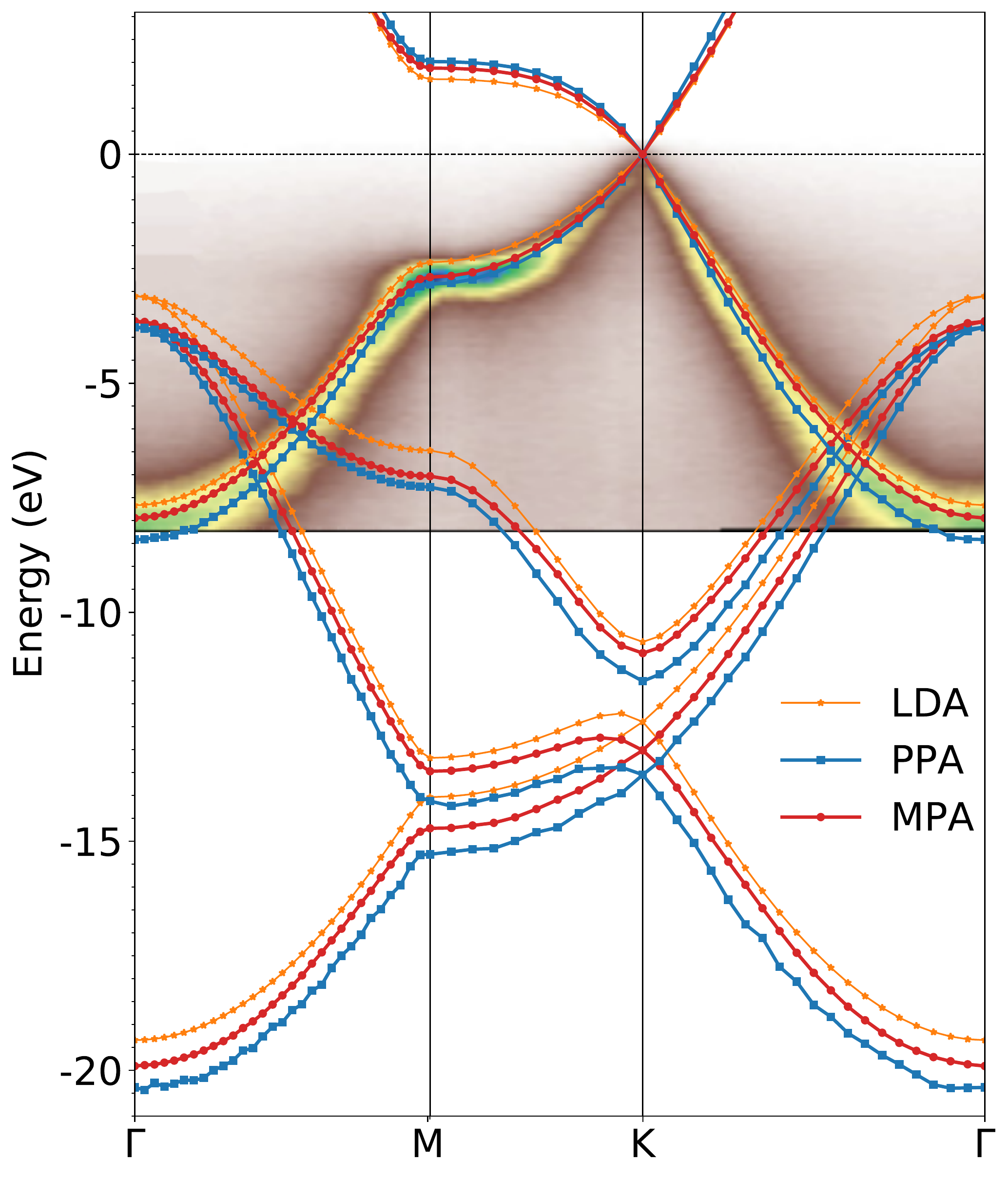}
\caption{Graphene QP band structure computed within DFT-LDA (orange), GW/PPA (blue) and GW/MPA (red). In the energy range from -8~eV up to the Fermi energy, we show the ARPES experimental measurements reported in Ref.~\onlinecite{Knox_2011}.
}
\label{Fig_exp}
\end{figure}

%============================
\subsection{Comparison with experiments and existing literature}\label{Sec_compare_exp}
%============================

In Fig.~\ref{Fig_exp}, we compare the DFT-LDA with the QP band structure obtained within PPA and MPA. The calculated occupied $\pi$ bands are compared with experimental ARPES measurements from Ref.~[\onlinecite{Knox_2011}].
Additional experimental data, though limited to the Dirac cone region, can be found in Ref.~[\onlinecite{Bostwick_2007}].
As expected, DFT fails in describing key features of the graphene band structure. In particular, it underestimates the Fermi velocity, the gap at $M$ and the bandwidth of the $\pi$ band.
While PPA improves on the first two features, yielding a reasonable QP band structure near the Fermi energy, its accuracy diminishes for deeper states, for which it tends to overestimate the QP corrections. Conversely, MPA provides a more comprehensive description of the band structure over a broader energy range, in agreement with ARPES results.

In Tab.~\ref{Tab_QP}, we report the converged values of the Fermi velocity and gap at $M$ obtained with  PPA and MPA, with a grid of $120\times 120$ $\k$-points. DFT results are also shown for completeness. For the Fermi velocity, the experimental measure from Ref.~[\onlinecite{Knox_2011}] is included both in Tab.~\ref{Tab_QP} and in the bottom panel of Fig.~\ref{Fig_gap}, with the error bar shown in gray.
The Fermi velocity measured in Ref.~[\onlinecite{siegel2011many}] is within the error bar of Ref.~[\onlinecite{Knox_2011}].
The Fermi velocity  computed within PPA is 1.18$\times10^6$ m/s, consistent with previously reported $GW$ values, which range from 1.12 to 1.25(5)$\times10^6$~m/s. More specifically, 1.12$\times10^6$~m/s was reported from both PPA and the contour deformation methods in Ref.~[\onlinecite{trevisanutto2008ab}], 1.15$\times10^6$~m/s  from PPA~\cite{Yang2009} and 1.25(5)$\times10^6$~m/s again from PPA.~\cite{Attaccalite_2009}
Within MPA, the Fermi velocity is around 1.1$\times 10^6$~m/s, smaller than our PPA result, so that both approximations yield a very good agreement with the experiment.~\cite{Knox_2011}

Significant differences between PPA and MPA are observed in the estimation of the QP gap at the M point, with PPA providing a larger correction to the DFT eigenvalues. Specifically, the difference between PPA and MPA at the $M$ point is approximately 0.34~eV, which is notably larger than the differences reported for quasi-particles in semiconductors and simple metals.~\cite{Valido_2021,leon2023efficient} As in the case of more complex metals such as Cu,~\cite{leon2023efficient} this can be explained by the improved description of the low-energy features of the screened interaction, which are relevant in the case of graphene.
As expected from other full frequency results,~\cite{Miglio2012EPJB} the difference between the MPA and PPA descriptions builds up for the states further away from the Fermi energy. For example, the state near the $\Gamma$ point at around $-8$~eV, is over-corrected by PPA by $0.47$~eV with respect to MPA. This can be indeed ascribed to the incomplete description of the self-energy structures below -4~eV (see Fig.~\ref{Fig_SE}), more important for deeper states. 
Moreover, as shown in Fig.~\ref{Fig_exp}, other representations issues of the PPA self-energy are apparent in the  noise of the bands corresponding to quasi-particles between -14 and -21~eV.

The QP band structure obtained from MPA is, in general, in good agreement with experimental measurements in the whole considered energy range, even if, in the present calculations, electron-phonon (e-ph) coupling and vertex corrections are not included, and no self-consistency is considered.
The e-ph coupling reduces the Fermi velocity by around $4\%$ with respect to the DFT value,~\cite{park2007velocity,park2008electron} 
an amount smaller than the difference between PPA and MPA values.
The inclusion e-ph coupling would be important instead if we were to calculate QP lifetimes,~\cite{park2009angle,PhysRevLett.102.076803} which are however out of the scope of this work.
To the best of our knowledge, neither the effect of self-consistency nor the inclusion of vertex corrections have been investigated beyond the Dirac Hamiltonian model.~\cite{Abrikosov_1970,Gonzales_1994,Gonzales_1999,Basko_2008}

%============================
\subsection{Velocity renormalization at the Dirac cone}\label{Sec_dirac_cone}
%============================

In this Section, we examine the band structure of graphene in proximity to the Dirac point, in particular we aim at elucidating %on the rendition of 
the logarithmic behavior of the $\pi$ band and the divergence of the Fermi velocity.~\cite{elias_dirac_2011,siegel2011many}
For that purpose, we have performed a series of calculations using the PPA and MPA@Wav methods. 
We define a $k$-dependent electron velocity $\gamma (k) \equiv d \varepsilon(k)/dk$, with $k$ arbitrarily chosen along the $\Gamma K$ direction of a $K$ centred $\mathbf{k}$-grid, i.e. $k=0$ corresponds to $K$, and consequently, the Fermi velocity can be expressed as $\gamma (k=0)$.

The energy variations in the vicinity of $K$ are abrupt enough to hinder the accurate description of the band with a finite number of $\mathbf{k}$-points.
To circumvent this difficulty, we have designed a hyperbolic model of the QP energies near the
Dirac cone based on the derivation presented in App.~\ref{apendix_hf}. According to this model, the energy dispersion is described by the expression: 
\begin{equation}\label{eq:E_sm_model}
\begin{split}
    \varepsilon_D(k) &=\gamma_c k + f \gamma_c \frac{k}{2} \left[\cosh^{-1}{\left(\frac{2 k_c}{k + k_s}\right)} +\frac{1}{2}\right] %\\
    %\gamma_D(k, k_s) &\equiv \frac{d \varepsilon_D(k, k_s)}{d k},    
\end{split}
\end{equation}
where $k_c$ is the ultraviolet cutoff fixed to the boundary of the Brillouin zone, here $k_c=1/2.15~$\AA$^{-1}$,~\cite{siegel2011many,Attaccalite_2009} and $k_s$ accounts for the smearing of the band occupations  close to the Fermi energy (see below).
In the above expression, $\gamma_c$ and $f$ are parameters of the model related, respectively, to the electronic velocity for large $k$, and to the (dimensionless) weight of the electron-electron interactions.

Models previously used to fit experimental data~\cite{siegel2011many,elias_dirac_2011} and $GW$ calculations,~\cite{Trevisanutto_2008,Attaccalite_2009} take into account only the first terms of a Taylor expansion around $k=0$.
The present model, as detailed in App.~\ref{apendix_hf}, has the same Taylor expansion of the exact Hartree-Fock solution of the Dirac Hamiltonian, at all orders for $k \to 0$. Moreover, it follows very closely the exact solution for the whole $k$ range (see Fig.~\ref{fig:cone_model}), with a maximum deviation, for $k=k_c$, of about 0.8\% for what concerns the nonlinear part and less than 0.1\% if we consider $\varepsilon_{D}(k)$. 

As explained in Sec.~\ref{Sec_self_energy}, $GW$ calculations with discrete grids present a discontinuity at the Dirac point, hindering the accurate description of the Dirac cone. It is common to address this issue by making use of a Fermi-Dirac distribution with a small smearing $s$, compatible with the distance between the points of the discretized $\mathbf{k}$-grid.~\cite{trevisanutto2008ab,Attaccalite_2009} The effect of $s$ is included in Eq.~\eqref{eq:E_sm_model} by shifting  the $k$ points by $k_s$, modelled through the linear dependence, $k_s = s/\gamma_c$. The presence of a finite smearing preserves the logarithmic dependence of the Fermi velocity but removes the divergence at $K$, similarly to what happens in graphene with doping.~\cite{trevisanutto2008ab,Attaccalite_2009,Elias_2011}
The dependence of the parameters of the model on the value of the smearing is analysed in detail in the \suppinfo.

\begin{figure}%[hbtp]
\includegraphics[width=0.48 \textwidth]{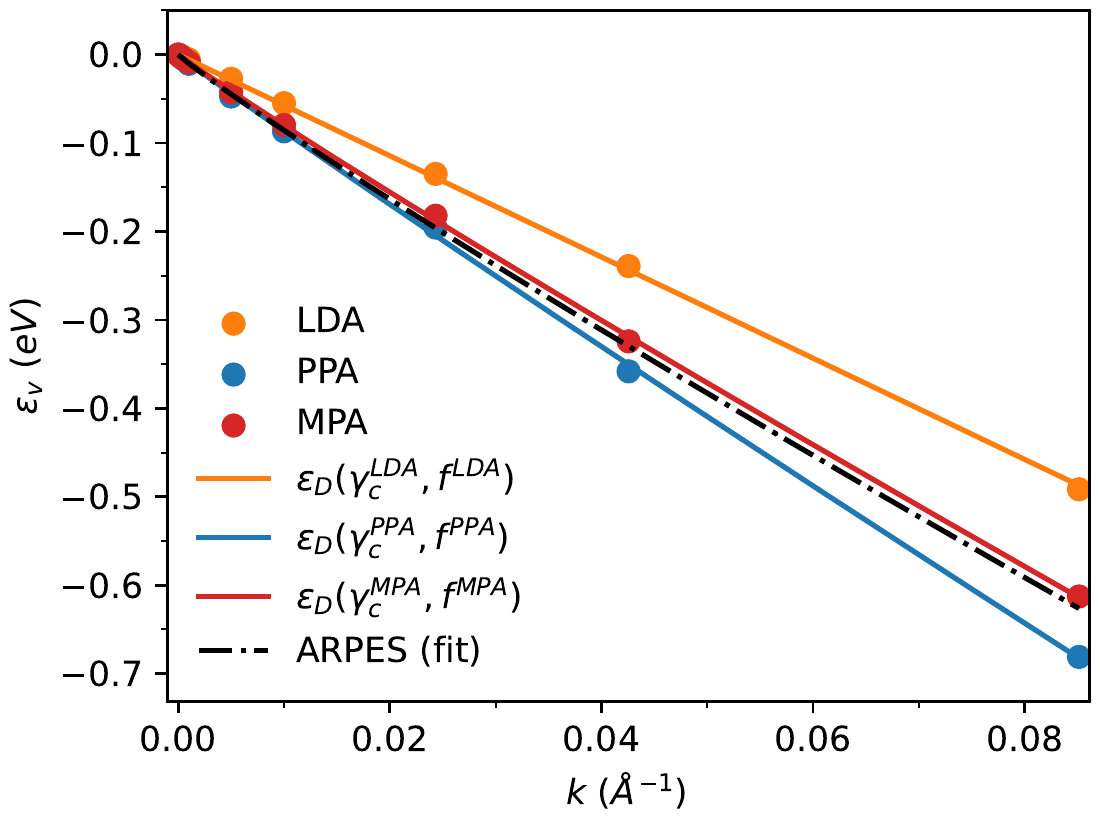}
\caption{Graphene valence $\pi$ band in the region close to the  Dirac point ($k=0$), computed with DFT-LDA, $GW$-PPA@Wav and $GW$-MPA@Wav, with a 120$\times$120 $\mathbf{k}$-grid plus 6 extra points (see Ref.~[\onlinecite{Attaccalite_2009}]) and a Fermi-Dirac smearing of $0.0136$~eV ($1$~mRy) for the $GW$ calculations. The lines show the fit of the numerical values using the hyperbolic model of Eq.~\eqref{eq:E_sm_model}. We compare with a logarithmic model fitted to ARPES measurements from Ref.~[\onlinecite{siegel2011many}]. The values of the fitted parameters are shown in Tab.~\ref{tab:model_par}.
\label{fig:PPA-MPA-model}
}
\end{figure}

In Fig.~\ref{fig:PPA-MPA-model} we compare the $\pi$-band of graphene close to the Dirac point, computed at different levels of theory, with the model in Eq.~\eqref{eq:E_sm_model} parameterized on {\it ab initio} data. In Tab.~\ref{tab:model_par} we summarise the parameters of the model. 
We also report the electronic velocity in the linear regime, 
%$\gamma_L$, computed numerically as 
$\gamma_L = \varepsilon_D(k_L)/k_L$ at $k_L=0.085$~\AA$^{-1}$, for a model without smearing, i.e. $k_s = 0$. The calculations were done within DFT-LDA, $GW$-PPA and $GW$-MPA@W-av, considering a $120\times120$ $\mathbf{k}$-grid, complemented with an additional set of points around $K$, as shown in Fig.~\ref{fig:PPA-MPA-model}, and described in Sec.~\ref{Sec_comp_details}. The adopted smearing value is  $s=0.0136$~eV (see \suppinfo).  

The model follows very closely the numerical values for the three approaches, and the computed values of $\gamma_L$ are consistent with the ones reported in Tab.~\ref{Tab_QP}, with differences  below $5$\% and $3$\% for PPA and MPA respectively.
Regarding instead the asymptotic velocity in the model, $\gamma_c$,
$GW$ calculations with PPA shows an increase with respect to DFT while MPA presents a similar value, which is also close to the value fitted with a logarithmic model to ARPES measurements from Ref.~[\onlinecite{siegel2011many}].

As expected, for LDA the fitted model has a zero $f$, whereas for both $GW$ calculations $f$ is finite. As a result of the screening, the $GW$ values are smaller than the Hartree-Fock value estimated for the fine structure constant of graphene from empirical models, $f^\text{HF}=\alpha /4\simeq0.55$ with $\alpha=2.2$.~\cite{Reed2010Science} 
If we use the model derived in Ref.~[\onlinecite{elias_dirac_2011}] for the RPA screening, consistent with our Eq.~\eqref{Eq_chi_0_D_stat}, and relate the fine structure constant with the parameter $f$, we obtain $f=\frac{\alpha}{4(1+\pi\alpha/2)} \simeq 0.12$, which is very close the present $GW$ values, $f^{\text{PPA}}=0.10$ and $f^{\text{MPA}}=0.15$.
In Refs.~[\onlinecite{Gan2016PRB,Reed2010Science}], the influence of screening is estimated, from inelastic x-ray scattering on graphite, to renormalize the effective fine structure constant of graphene giving $\alpha^*\simeq 0.25\text{-}0.35$, while the fit of ARPES measurements from Ref.~[\onlinecite{siegel2011many}] results in $\alpha^* = 0.40 \pm 0.01$ and therefore $f=\alpha^*/4\simeq0.1$, again consistent with the $GW$ values found here for $f$.
Among the three approaches, the MPA Dirac cone, with its larger $f$, is the one that most deviates from the linear behavior, revealing the importance of the dynamical nature of $W$, that is accounted for only in a simplified way within PPA.

\begin{table}[]
    \centering
\begin{ruledtabular}    
    \begin{tabular}{c|ccc}
            & $\gamma_c$~($10^6$~m/s) & $f$  & $\gamma_L$~($10^6$~m/s)\\[5pt]
\hline\\[-3pt]
DFT-LDA&  $0.87 \pm 0.005$  &  0 &  $\gamma_c$        \\
$GW$-PPA &  $1.04 \pm 0.06$  &  $0.10 \pm 0.04$ &  1.24        \\
$GW$-MPA &  $0.87 \pm 0.04$  &  $0.15 \pm 0.03$ &  1.12  \\
Exp~\cite{siegel2011many} &  $0.86 \pm 0.02$  &  $0.10 \pm 0.002$&  1.16
    \end{tabular}
\end{ruledtabular}
    \caption{Parameters of the hyperbolic model of Eq.~\eqref{eq:E_sm_model} as fitted to the numerical values of the DFT, $GW$-PPA and $GW$-MPA calculations of the graphene $\pi$ band close to the Dirac point. The parameters of a logarithmic model fitted to ARPES measurements from Ref.~[\onlinecite{siegel2011many}] are also included. To compare with other results in the literature, we also report $\gamma_L$, the electronic velocity in the linear regime computed numerically as $\gamma_L = \varepsilon_D(k_L)/k_L$, where $k_L=0.085$~\AA$^{-1}$.
    The same procedure is use to compute $\gamma_L$ with the logarithmic model from Ref.~[\onlinecite{siegel2011many}].
Notice that all the reported velocities correspond to their absolute values.
    }
    \label{tab:model_par}
\end{table}

\begin{figure}%[hbtp]
\includegraphics[width=0.49 \textwidth]{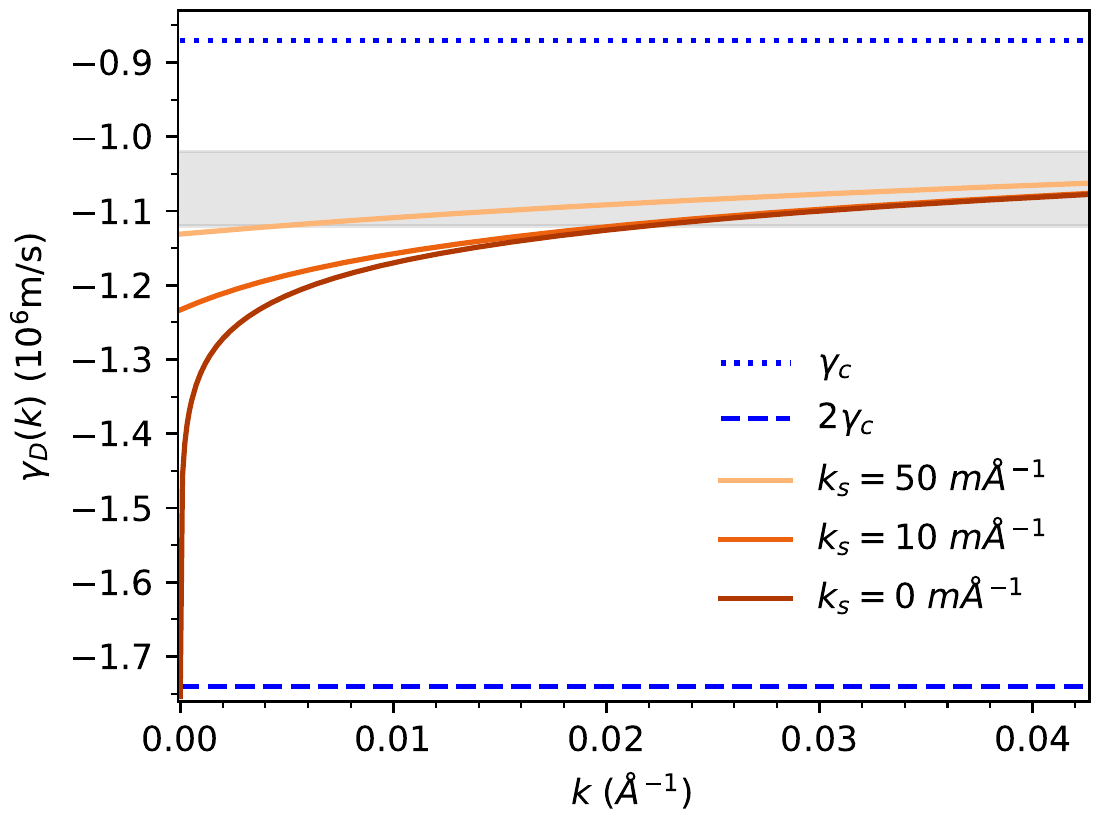}
\caption{Momentum dependent electronic velocity of the graphene band close to the Dirac point computed as the derivative of the model in Eq.~\eqref{eq:E_sm_model}, evaluated with the MPA@W-av parameters listed in Tab.~\ref{tab:model_par} for different values of the Fermi-Dirac broadening. As in Fig.~\ref{Fig_gap}, the gray bar corresponds to the experimental value of the velocity in the linear regime and its uncertainty, measured in Ref.~[\onlinecite{Knox_2011}].
%\AGnote{I found confusing that the yticks are incremented by $0.15$. Is it possible to use $0.10$ or $0.20$? (Maybe if $0.20$ with more dense secondary ticks).}
\label{fig:v-model}
}
\end{figure}

The small value of $f$, even in MPA, elucidates why the exact behavior of the velocity close to $K$ is difficult to assess using solely numerical results obtained from a discrete $\mathbf{k}$-grid. 
In Fig.~\ref{fig:v-model} we show the velocity dispersion derived from Eq.~\eqref{eq:E_sm_model}, with the parameters fitted to the MPA results for the valence $\pi$ band, with  different $k_s$ values. As a reference, we represent also two extra lines corresponding to a linear dispersion with $\gamma=\gamma_c$ and $\gamma=2 \gamma_c$.
The logarithmic renormalization of the velocity is well described, including the case $k_s=0$, where the divergence is restored. Only in a very small vicinity of $K$ and for $k_s=0$, there is a significant change in the velocity:
it reaches $2 \gamma_c$ for a  distance from $K$  of 
approximately  $k=1.3\times10^{-6}~$\AA$^{-1}$, 
which corresponds to a $\mathbf{k}$-points Monkhorst-Pack grid of the order of $10^6 \times 10^6$.
Obtaining a reliable $\gamma (k)$ from the QP energies using finite differences would require a very large number of points computed in very close proximity to $K$ and, in practice, no smearing.
This also suggests that, in order to capture experimentally the doubling of the electronic velocity, one would need a graphene sample with a size exceeding $10^{-4}$ m.

\section{Conclusions}
\label{Sec_conclusions}

In this work we propose a method that combines together the MPA~\cite{Valido_2021} approach and the W-av~\cite{Guandalini2023npjCM} scheme (labelled MPA@W-av here), aimed at efficiently computing the $G_0W_0$ self-energy with coarse-grained frequency and BZ samplings. 
First we provide and discuss the theoretical formulation of the approach, which has been implemented in the \yambo{} package. 
Then, in order to illustrate the method, we apply it to the calculation of the self-energy and QP band structure of graphene.

We show that MPA@W-av describes very accurately the sharp dependence of the dynamic inverse dielectric function on the momentum transfer.
The comparison with PPA reveals the importance of the accurate description of the frequency dependence of the screening provided by MPA.
W-av accelerates the convergence of both real and imaginary parts of $\Sigma$, in particular in the frequency region around the solution of the QP equation.
This affects the QP corrections, reducing, for example, the MPA gap at $M$ and improving the agreement with the experimental values.
The comparison with ARPES measurements shows that, while the PPA description is qualitatively accurate for energies close to the Fermi energy, MPA extends the agreement to a significantly larger energy range.

Moreover, given the accuracy of the approach,
we have also addressed the electronic velocity renormalization at the Dirac point, focusing in particular on the logarithmic behavior approaching the Dirac point. Our findings show that this behavior is difficult to render with a finite set of quasi-particle energies on a discrete $\k$-grid. To circumvent this difficulty we have designed a simple but very accurate model, that reproduces very accurately the exact solution of the Hartree-Fock self-energy for a Dirac Hamiltonian, then generalized to account for screening effects  at the $GW$ level. The model was used to fit the sets of numerical results in order to give an analytical description of the Dirac cone and of the renormalization of the Fermi velocity.

Overall, the proposed MPA@W-av method combines the efficiency and the accuracy gains already delivered by the separate implementation of MPA and W-av. It provides $GW$ converged results using $\mathbf{k}$-point grids not much denser than those needed for DFT calculations, as previously reported for semiconductor using PPA,~\cite{Guandalini2023npjCM} while providing an accuracy similar to full-frequency methods, but using around few tens of frequency sampling points (20 here), instead of hundreds or thousands.~\cite{Valido_2021,leon2023efficient}
Finally, we would like to emphasise that MPA@W-av can be easily used to treat systems with different dimensionalities (1D and 3D) and with different screening properties (such as metals), with the potential to become a widespread tool within the $GW$ methodology.

%========================
\section*{Acknowledgments}
%========================
%
We acknowledge stimulating discussions with Francesco Mauri, Paolo Barone, Savio Laricchia, Miki Bonacci, Simone Vacondio, Matteo Zanfrognini,  and Giacomo Sesti. This work was partially supported by 
MaX -- MAterials design at the eXascale -- a European Centre of Excellence funded by the European Union's program 
HORIZON-EUROHPC-JU-2021-COE-01 (Grant No. 101093374), 
ICSC – Centro Nazionale di Ricerca in High Performance Computing, Big Data and Quantum Computing, funded by European Union –NextGenerationEU - PNRR, Missione 4 Componente 2 Investimento 1.4.
%, and the Italian national program PRIN2017 2017BZPKSZ “Excitonic insulator in two-dimensional long-range interacting systems”.
We acknowledge CINECA for computational resources, awarded via the ISCRA program.

\appendix
%%%%%%%%%%%%%%%%%%%%%%%%%%%%%
\section {Graphene Dirac model for the polarizability}\label{Sec_Dirac_term}
%%%%%%%%%%%%%%%%%%%%%%%%%%%%%

As detailed in Sec.~\ref{Sec_Wav}, the W-av method uses an auxiliary function, $f(\q)$, to interpolate $\scrc$ 
between nearest neighbour points within the BZ. In practice, it is convenient to tailor
the analytical form of this function to the behavior of $W$ in the region of small $\bf q$ vectors. For 2D semiconductors, for which W-av was originally formulated,~\cite{Guandalini2023npjCM} $f(\q)$ depends quadratically on $\q$ in this regime.~\cite{Rasmussen_2016} However, freestanding graphene has a semimetallic behavior at the Dirac cone and, in the static limit ($\o\to0$) and for  $\q\to0$, the $f_{\G=\G'=0}(\q)$  element depends linearly on $\q$, thus requiring a different interpolating function. 

Moreover, in the $\q \to 0$ limit, the contributions from vanishing transition energies, such as intra-band transitions in metals, are not trivially included in the calculation of the polarizability.
When neglected, a spurious gap opens, which dramatically slows  the convergence with respect to the BZ sampling.~\cite{Cazzaniga_2008} 
For metallic systems, several computational schemes have been developed to overcome the problem, e.g. computationally intensive Fermi-surface integrations,~\cite{Maksimov_1988,Lee_1994,Marini_2001_2} sub-sampling methods~\cite{daJornada_2017} or fitting schemes over the BZ.~\cite{Cazzaniga_2008} A semi-empirical alternative is the inclusion of a Drude correction in the irreducible polarizability.~\cite{Marini_2001_2}
In graphene, vanishing transition energies occur near the Dirac point $K$.
In the following, we show how this contribution can be taken into account by considering
 the long-wavelength ($\q \to 0$) contribution of the Dirac cone transitions to $\chi_0$ as based on a Dirac Hamiltonian.~\cite{Neto_2009}

 The 2D macroscopic  independent-particle polarizability $\chi^0$ reads:
\begin{multline}
    \chi_{\G=\G'=0}^0(\q,\o) =
    2\sum\limits_{v,c}
    \int \frac{d\k}{(2\pi)^2}
    |\dm_{cv}(\k,\q,\G=0)|^2 \times \\
    \left[
    \frac{1}{\o-(\eig_{c\k}-\eig_{v\k-\q})+i\eta}
    -\frac{1}{\o+(\eig_{c\k}-\eig_{v\k-\q})-i\eta}
    \right] ,
    \label{Eq_chi_0}
\end{multline}
where $c$ and $v$ refer to conduction and valence band indexes, the limit $\eta \to 0^+$ is implicit as in Eq.~\eqref{eq_GW_expl}, and the factor $2$ accounts for the spin degeneracy. 
For metals or semimetals, as $\o$ and ${\q}$ approach zero, the term in square brackets diverges due to the simultaneous vanishing of the frequency $\o$, and the energy difference $\eig_{c\k}-\eig_{v\k-\q}$. However, since the matrix elements  $\dm_{cv}(\k,\q,\G=0)$ also vanish for $\q\to0$, this leads  to numerical difficulties when computing
$\chi_{\G=\G'=0}^0(\q,\o)$ in these limits.

In the case of graphene, we split the integration over the BZ in Eq.~\eqref{Eq_chi_0} into two intervals, $\k \in D_{K}$ and $\k \not\in D_{K}$, where $D_{K}$ is a small circular region around the Dirac cone centred at $K$. For sufficiently small $D_{K}$, only the $\pi/\pi^*$ bands contribute to the polarizability, from here on denoted as $v/c$. We can then write the long wavelength contribution to the polarizability  in this region as:
\begin{multline}
    \chi^0_{D}(\omega) \equiv \lim\limits_{\q \to 0} 4\int\limits_{D_K} \frac{d\k}{(2\pi)^2}
    |\dm_{{ v}{c}}(\k,\q)|^2\times\\
    \left[
    \frac{1}{\o-(\eig_{{c}\k}-\eig_{{v}\k-\q})+i\eta}\right.
    \left. - \frac{1}{\o+(\eig_{{c}\k}-\eig_{{v}\k-\q})-i\eta}
    \right],
    \label{Eq_chi_0_D}
    \end{multline}
where the factor $4$ accounts for both the spin and valley degeneracy.  

As an alternative to numerical integration, Eq.~\eqref{Eq_chi_0_D} can be evaluated analytically by considering a Dirac model of the cone, with 
band dispersion and  square modulus of matrix elements $\dm_{{v}{c}}(\k,\q)$ given by:
\begin{equation}
\begin{split}
   & \eig_{c/v\k} = \pm\gamma|\k|, \\
   & |\dm_{\pm}(\k,\q)|^2 = \frac{1}{2}\left[ 1 \pm \cos(\theta_{\k,\k-\q}) \right] ,
    \label{Eq_eig_D}
\end{split}
\end{equation}
where $\rho_{vv} = \rho_{+}$ and $ \rho_{vc}=\rho_{-}$, $\mathbf{k}=0$ now corresponds to the Dirac point $K$, $\gamma$ is the Fermi velocity, and $\theta_{\k,\k-\q}$ the angle between the vectors $\k$ and $\k-\q$.
By inserting Eqs.~\eqref{Eq_eig_D} into Eq.~\eqref{Eq_chi_0_D}, the integral in Eq.~\eqref{Eq_chi_0_D} is straightforwardly calculated in elliptic coordinates.
After taking the limits $\eta \to 0^+$ and $\q \to 0$, we find:
\begin{equation}
    \chi^0_{D}(\omega) = -\frac{q^2}{4}\left[
    \frac{\Theta(\gamma q-\o)}{\sqrt{\gamma^2q^2-\o^2}}+
    i\frac{\Theta(\o-\gamma q)}{\sqrt{\o^2-\gamma^2q^2}}
    \right],
    \label{Eq_chi_0_D_dyn}
\end{equation}
where $\Theta$ is the Heaviside step function. Finally, we evaluate Eq.~\eqref{Eq_chi_0_D_dyn} in the static limit as:
\begin{equation}
    \chi^0_{D}(\omega=0) = -\frac{q}{4\gamma} \ ,
    \label{Eq_chi_0_D_stat}
\end{equation}
which allows one to compute the Dirac cone contribution to the static-independent particle polarizability.
We note that this expression was used in Refs.~[\onlinecite{Shung_1986,Gorbar_2002,Ando_2006,Wunsch_2006,Barlas_2007,Wang_2007,Sarma_2007}], for all values of momentum transfer $\q$ in the whole BZ, while, in our approach, Eq.~\eqref{Eq_chi_0_D_dyn} is considered only in the long wavelength limit 
and within the $D_K$ region. Otherwise, the polarizability is calculated fully {\it ab-initio} via Eq.~\eqref{Eq_chi_0}.

In Sec.~\ref{Sec_self_energy} we have discussed the advantages of introducing the correction $ \chi^0_{D}$ in the polarizability, as it accelerates the convergence of the quasi-particle calculation with respect to the number of {\bf k}-points.
Moreover, Eq.~\eqref{Eq_chi_0_D_stat} may be used to describe the Dirac cone contribution to the polarizability in other 2D Dirac semimetals and it can be generalized to different Dirac cone occupations, as in doped graphene.

\newcommand\deco[0]{%
  \par\vspace{1ex}
  \begin{center}
  {* * *}
  \end{center}
  \vspace*{1ex}\par
}

\section{Graphene Dirac Model for the self-energy}\label{apendix_hf}
Following Ref.~[\onlinecite{siegel2011many}], and in analogy with the quasi-particle equation, Eq.~\eqref{eq_QP_def}, the QP dispersion
of the occupied $\pi$ band of graphene can be written in atomic units as:
\begin{equation}
    \eig_{v\k}^{\mathrm{QP}} = \eig_{v\k} + {2 \gamma_c f} \int \frac{d \q}{(2\pi)^2} v(\q) | \rho_{vv}(\k, \q) |^2,
    \label{eq_QP_Dirac}
\end{equation}
where we have used the Dirac model of Eq.~\eqref{Eq_eig_D} redefining the Fermi velocity as $\gamma_c$. The factor $2$ accounts for the spin degeneracy and the Coulomb potential is given by $v(\q)=2\pi/q$. The integral corresponds to the Hartree-Fock (HF) self-energy $\Sigma^x_{v \k}$, 
rescaled by the factor $f$, introduced to account for screening effects, such as those arising from the electronic screening within the $GW$ approximation and/or the dielectric environment~\cite{Barnes2014PRB}. Therefore, $f = 1/\gamma_c$ in the case of the HF approximation when applied to suspended graphene. In practice, $\gamma_c$ and $f$ are used as free parameters when fitting DFT and $GW$ numerical results.

Here we focus of the integration of the rescaled HF term, which modifies the linear dispersion of the $v$ bands around the Dirac point, set at $\k=0$.
The integration is performed up to a maximum radial cutoff value $k_c$,~\cite{elias_dirac_2011,Barnes2014PRB} the ultraviolet limit within the framework of the renormalization group theory.~\cite{Barnes2014PRB}
We then rescale all the momentum vectors by $k_c$, $\q' \equiv \q/k_c$ and $\k' \equiv \k/k_c$, and choose a coordinate system $\q'=(q_x,q_y)$ in which $\k'$ has only one component, $\k'=(k,0)$.
Substituting Eq.~\eqref{Eq_eig_D} into Eq.~\eqref{eq_QP_Dirac},
the rescaled self-energy, $\Sigma'_{v k} \equiv \Sigma_{v k}{/k_c}$, in Cartesian coordinates reads:
\begin{equation}
    \Sigma'_{v k} =- \frac{{\gamma_c f}}{2\pi} \int \frac{d \q'}{|\q'|} \left(1- \frac{k - q_x}{\sqrt{(k - q_x)^2 + q_y^2}} \right),
    \label{eq:Sx_cartesian}
\end{equation}
where the integration domain is given by $0\leq q_x^2+q_y^2\leq1$.

In Eq.~\eqref{eq:Sx_cartesian} only $k$ dependent terms are relevant, since the other term will only change the energy reference for the vacuum level.~\cite{Barnes2014PRB}
Using polar coordinates, the integral of interest can be written as:
\begin{equation}
    \Sigma'_{v k} = -\frac{\gamma_c f}{2\pi} \int_0^{2\pi} d\theta \int_0^1 d r \frac{k - r \cos{\theta}}{\sqrt{k^2-2 k r\cos{\theta}+r^2}}.
    \label{eq:Sx_polar}
\end{equation}
We present here a slightly different solution with respect to the one in Refs.~[\onlinecite{hwang_density_2007,Pavlyukh2020PRB}], stressing the importance of higher order terms of its Taylor expansion. %\DVnote{cosi sembra che in Das Sarma c'è la stessa espessione in polar coord, ma non mi sembra sia così, forse questa frase va inserita prima?. In generale, per quello che viene dopo non è chiarissimo cosa è preso dalla letteratura e cosa è originale, forse va specificato.}. \DALnote{nella seconda Ref c'e la stessa  integrale (Eq. (68))}
Working out the radial integral, we get:

\begin{eqnarray}
   %\begin{split}
   \nonumber
    \Sigma'_{v k} &=& \frac{{\gamma_c f}}{2\pi} \int_0^{2\pi} d\theta \,
    \bigg\{ \cos{\theta}\left(k - \sqrt{k^2-2 k \cos{\theta}+1}\right) + \\ \nonumber 
    && \quad +k \sin^2{\theta} \bigg[ \tanh^{-1}(\cos{\theta}) + %\right.
    \\ %\nonumber
    && \left. \left. \qquad +\tanh^{-1} \left(\frac{1-k \cos{\theta}}{\sqrt{k^2-2 k \cos{\theta}+1}} \right) \right] \right\}.
    %\end{split}
    \label{eq:Sx_polar1}
\end{eqnarray}
Even if the angular integral cannot be solved exactly in terms of elemental functions, it can however be evaluated numerically up to any desirable precision. In order to arrive to an approximated analytical expression we derive the Taylor expansion of Eq.~\eqref{eq:Sx_polar1}, both for small and large $k$.

We consider the momentum-dependent electronic velocity, $\gamma(k) \equiv d\varepsilon/dk$, so that we can deal with the hyperbolic inverse tangents and get
\begin{equation}
    \frac{d \Sigma'_{v k}}{d k} = \frac{1}{k} \left[ \Sigma'_{v k} + \frac{{\gamma_c f}}{2\pi} \int_0^{2\pi} d\theta
    \frac{k - \cos{\theta}}{\sqrt{k^2-2 k\cos{\theta}+1}} \right].
    \label{eq:Sx_polar2}
\end{equation}
The remaining integral can be expressed in terms of complete elliptic integrals:
\begin{equation}
    \begin{split}
    &\int_0^{2\pi} d\theta
    \frac{k - \cos{\theta}}{\sqrt{k^2-2 k\cos{\theta}+1}} = \\
    &\frac{2}{k} \left[ (k+1) E_K \left(\frac{-4 k}{(k-1)^2}\right) +(k-1) E_E \left(\frac{-4 k}{(k-1)^2}\right) \right],
    \end{split}
    \label{eq:Sx_polar3}
\end{equation}
where, for a value $m<1$, the complete elliptic integrals of type 1 and type 2 are defined respectively as:
\begin{equation}
    \begin{split}
    E_K(m) & \equiv \int_0^{\pi/2} d \theta (1-m \sin^2{\theta})^{-1/2}, \\
    E_E(m) & \equiv \int_0^{\pi/2} d \theta (1-m \sin^2{\theta})^{1/2}.
    \end{split}
    \label{eq:elliptic_EK}
\end{equation}
The series expansion of $E_K(m)$ and $E_E(m)$ are known. If we consider the region of small $k$ values around the Dirac cone, we get
\begin{equation}
    \Sigma'_{v}( k\ll 1) = {\gamma_c f} \left[ -\frac{k}{2} \log{k} +c_1 k - \frac{k^3}{32} - \frac{3 k^5}{512} - O(k^7) \right],
    \label{eq:Sx_k_to_0}
\end{equation}
where $c_1=1/4+\log{2}$. The logarithm is the dominant term in the expansion, which diverges for $k\rightarrow0$,  a direct consequence of the finite integration cutoff, $k_c$.
On the other hand, if we consider the region close to the ultraviolet cutoff, $k \to 1^-$, we can define $u \equiv 1 - k$ and perform a Taylor expansion for $u \to 0^+$:
\begin{equation}
    \Sigma'_{v}(u \ll 1) = {\gamma_c f} \left[a_0 - a_1 u  + O(u^{2}) \right],
    \label{eq:Sx_k_to_1}
\end{equation}
where $a_0 = (2G+1)/\pi \approx 0.901432$, $a_1 = (2G-1)/\pi \approx 0.264812$ and $G$
is the Catalan's constant.

\begin{figure}
    \centering
    \includegraphics[width=0.40 \textwidth]{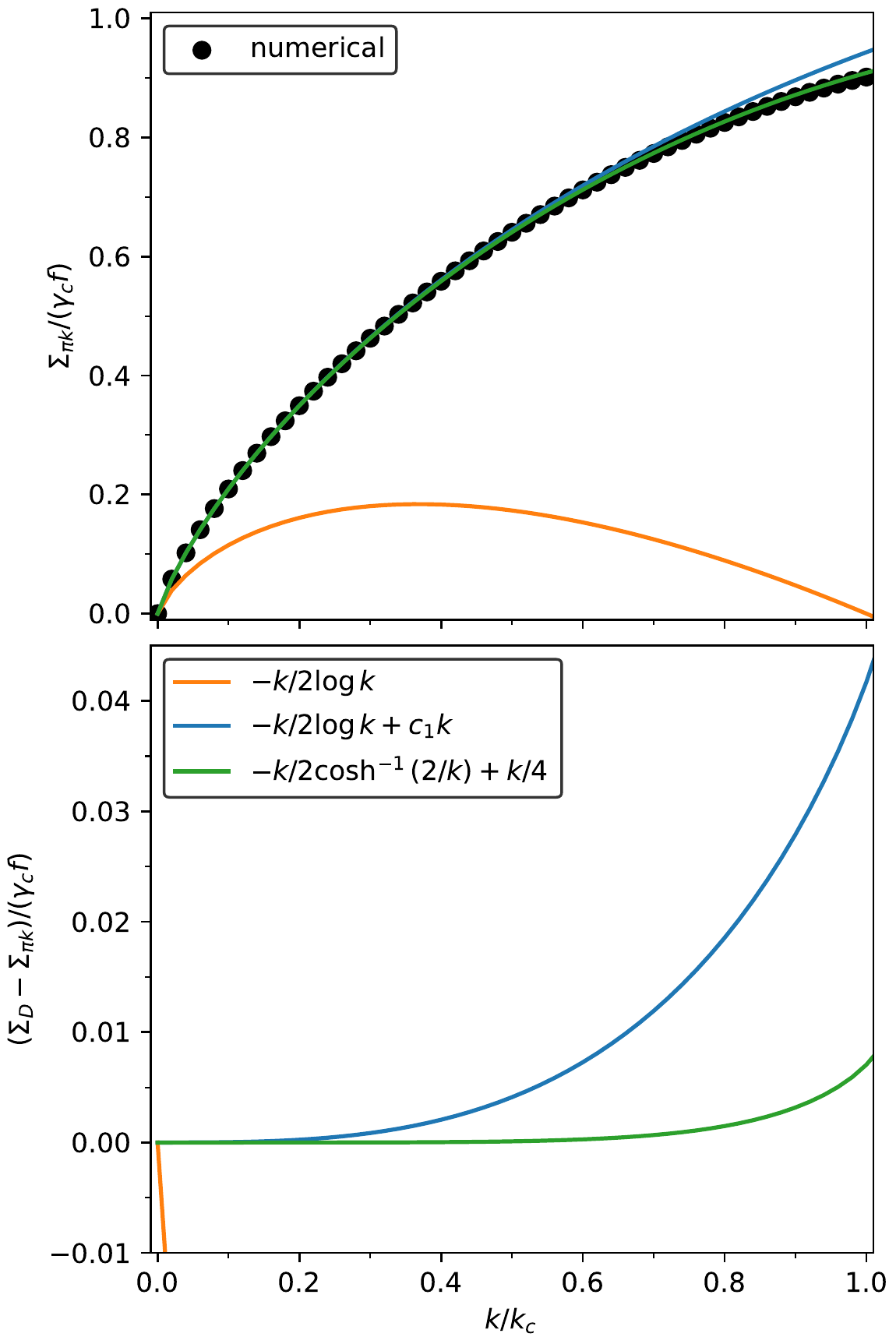}
    \caption{Top panel: comparison between $\Sigma'_{v k}$ obtained numerically from the exact expression given by Eq.~\eqref{eq:Sx_polar1} (black dots), the first (orange) and the first two (blue) terms of the Taylor expansion for $k=0$ in Eq.~\eqref{eq:Sx_k_to_0}, and the hyperbolic model in Eq.~\eqref{eq:sigma_D} (green). Bottom panel: difference between the models and the numerical solution.
    }
    \label{fig:cone_model}
\end{figure}

In order to reproduce both behaviors in Eqs.~\eqref{eq:Sx_k_to_0} and ~\eqref{eq:Sx_k_to_1} and better understand our numerical results, we propose the following analytical model:
%derived in the \suppinfo:
%
%Having in mind that the first term of the Taylor expansions of the inverse hyperbolic cosine is a logarithmic term that diverges for $k \to 0$, we have built a simple analytical model:
%
\begin{equation}
    \Sigma_D(k) = {\gamma_c f} \left[ \frac{k}{2} \cosh^{-1}{\left(\frac{2}{k}\right)} + \frac{k}{4} \right].
    \label{eq:sigma_D}
\end{equation}
Around $k=0$, $\Sigma_D(k)$ has the exact same Taylor expansion (up to infinite orders) of the one in Eq.~\eqref{eq:Sx_k_to_0}, while for $k \to 1^-$:
%it is similar to the one in Eq.~\eqref{eq:Sx_k_to_1}:
%
\begin{equation}
    \Sigma_D(u \ll 1) = {\gamma_c f} \left[a_0^1 - a_1^1 u + O(u^2) \right],
    \label{eq:model_k_to_1}
\end{equation}
with the coefficients, $a_0^1 = (1 + 2 \cosh^{-1}{2})/4 \approx 0.908479$
and $a_1^1 = (3 + \cosh^{-1}{2} - 4)/12 \approx 0.331129$, that are very similar to the ones in Eq.~\eqref{eq:Sx_k_to_1}.
%In fact, if we consider a non integer exponent $n =n_0 \approx 0.955327$, we get $a_0^{n_0} \approx a_0$ and $a_1^{n_0} \approx -0.310959$.
More insights about the construction of the model in Eq.~\eqref{eq:sigma_D} are provided in the \suppinfo.

In Fig.~\ref{fig:cone_model} we compare the results for the self-energy obtained numerically from the exact expression given by Eq.~\eqref{eq:Sx_polar1} (black), the first term (orange) and the first two terms (blue) of the Taylor expansion for $k=0$, with the model in Eq.~\eqref{eq:sigma_D}. In the bottom panel of Fig.~\ref{fig:cone_model} we show the difference between the numerical solution and the models. 
The models previously used to fit experimental data~\cite{siegel2011many,elias_dirac_2011} and $GW$ calculations~\cite{Trevisanutto_2008,Attaccalite_2009}, take into account only the first one or two terms of a Taylor expansion around $k=0$.
It can be seen that the logarithm follows the numerical solution only in the region very close to $k=0$. In contrast, accurate results are obtained in a much wider region when considering the first two terms of the expansion and the proposed model. 
The model with the hyperbolic cosine is particularly accurate, starting to deviate monotonically from the numerical result only above $k=0.5$ up to a maximum difference of about 0.8\% for $k=1$. We have used the model of Eq.~\eqref{eq:sigma_D}, with the additional linear term in Eq.~\eqref{eq_QP_Dirac}, $\eig_{{v}k} = - \gamma_c k$, to fit the $GW$ results presented in the main text.

\bibliography{bibliography}

\end{document}